\documentstyle[tighten,preprint,prd,aps]{revtex}
\begin{document}

\title{Scaling of the $\Upsilon$ spectrum in Lattice NRQCD}

\author{C. T. H. Davies\thanks{UKQCD collaboration}, A. Lidsey$^*$,
P. McCallum$^*$}
\address{University of Glasgow, Glasgow, G12~8QQ, UK}
\author{K. Hornbostel}
\address{Southern Methodist University, Dallas, TX 75275}
\author{G. P. Lepage}
\address{Newman Laboratory of Nuclear Studies, 
Cornell University, Ithaca, NY 14853}
\author{J. Shigemitsu}
\address{The Ohio State University, Columbus, OH 43210}
\author{J. Sloan}
\address{University of Kentucky, Lexington, KY 40506}

\maketitle

\begin{abstract}
We present results for the spectrum of $b\overline{b}$ bound states
in the quenched approximation for three different values of the
lattice spacing, in the range 0.05fm to 0.15fm. We find our
results for spin-independent splittings
in physical units to be independent of the lattice
spacing, indicating the absence of systematic errors from
discretisation effects.
Spin-dependent splittings are more
sensitive to the lattice spacing and higher order corrections
to the action; we discuss the size of these
effects and what can be done to arrive at a physical result.
\end{abstract}
\narrowtext

\section{Introduction}
Accurate calculations of the hadron spectrum in lattice QCD require
control of systematic errors. This has become a very important issue now that
statistical errors have been reduced in recent years to the point
where systematic errors can dominate the results reported.

A major source of systematic error is that arising from the use of
a space--time lattice with finite lattice spacing. All operators in
the continuum Lagrangian must be replaced with discrete versions and
discretisation
errors consequently appear. This means that physical results
(for example a mass in GeV)
depend
on the value of the lattice spacing. This is obviously wrong.
Since the lattice is simply a regulator for the theory,
physical results must not depend upon its value.
One approach has been to extrapolate
to zero lattice spacing for a `continuum' result. This is difficult
numerically, especially if the variation with lattice spacing is
severe. However, recent
progress in understanding discretisation errors and how to formulate
an improved action \cite{improve,wittig}, has
meant that we can obtain essentially continuum results at finite
values of the lattice spacing. The lattice spacing dependence is reduced
to such a low level that extrapolation is unnecessary.

The spectrum of bottomonium bound states is
one of the most accurate calculations that can be done on the
lattice \cite{oldups}.
Since the $b$ quarks are non-relativistic in these systems ($v^2/c^2 \approx
0.1$) a non-relativistic
action can be used \cite{gpl1,gpl2}. This allows a $b$ quark propagator to be
calculated on one sweep through the gluon field configuration
with low computational cost. Multiple
sources can be used on a single configuration because
the bound states are much smaller than the volume of a typical lattice.
Also sources for both ground and excited states can be used, allowing
multi-exponential fits to hadron correlators and improving the
confidence in the fitted masses.
These techniques mean that very small statistical errors can be obtained
and the improvement of systematic errors becomes a priority.
In this paper we discuss the issue of discretisation errors
for the bottomonium spectrum \cite{japan}.

The approach that we use, Non-relativistic QCD (NRQCD) \cite{gpl1,gpl2},
 is an effective field theory. Its
Lagrangian is suited to a description of non-relativistic quarks since
operators are classified according to the powers of $v^2/c^2$ they
contain, where $v$ is the velocity of the heavy quark. The number of
operators to be included can then be truncated at a fixed order in $v^2/c^2$
and this is clearly a sensible thing to do if $v^2/c^2 \ll 1$.
The renormalisability of QCD is lost in this process but
physical results are still obtained by putting an explicit momentum
cut-off into NRQCD. This cut-off should exclude relativistic
momenta and thus be of the same order, or smaller, than the heavy
quark mass. On the lattice this cut-off is provided by the
lattice spacing, with $Ma \stackrel{>}{\sim} 1$. The excluded momenta cause
renormalisation of the coefficients of the NRQCD operators
when, say, lattice NRQCD is matched to full continuum QCD.
The coefficients will be well-behaved and essentially cut-off
independent provided that the cut-off is not
too large.  Any attempt to take the cut-off to infinity (lattice
spacing to zero) will
cause them to diverge as the non-renormalisability of the theory
becomes apparent. Thus, no continuum extrapolation can be done for
lattice NRQCD. However, as discussed above, a continuum extrapolation is
not necessary for a suitably improved action.
All that is necessary is to demonstrate
lattice spacing independence of physical results.
For the bottomonium spectrum from NRQCD this should
be possible in a
region of lattice spacing $M_b a \stackrel{>}{\sim} 1$, and this is what
we show in this paper.

The size of discretisation errors will vary from one quantity to
another. In general it is to be expected that the coefficient of the
dependence on {\it a} should represent some typical momentum
scale appropriate to that quantity \cite{john}. For the light hadron
spectrum this would then be a few hundred MeV.  For heavy hadrons
the scale of discretisation errors is likely to be larger. The
scale is {\it not} set by the heavy quark mass since this is
an irrelevant scale to the dynamics of the bound states. It is
set rather by typical momenta exchanged inside the hadrons. For
bottomonium these momenta are of order 1 GeV and so discretisation
errors might be expected to present a problem on coarse lattices
if the action is not improved.
Here we report results with leading order ($a^2$) discretisation errors
removed from spin-independent terms, but not from the spin-dependent terms
(which are of lower order in the non-relativistic expansion).

Section 2 describes the lattice calculations and results at three different
values of the lattice spacing. Section 3 discusses
the scaling behaviour of spin-independent and spin-dependent splittings.
Section 4 contains our conclusions.

\section{NRQCD calculations and results}

Quark propagators in lattice NRQCD are determined, in a single pass
through the gauge-field configuration, from evolution equations that
specify the propagator for $t>0$ in terms of its value at $t=0$.
We use here: \cite{oldups,evolve}
\begin{eqnarray}
 G_1 &=&
  \left(1\!-\!\frac{aH_0}{2n}\right)^{n}
 U^\dagger_4
 \left(1\!-\!\frac{aH_0}{2n}\right)^{n} \, \delta_{\vec{x},0}  \nonumber \\
  G_{t+1} &=&
  \left(1\!-\!\frac{aH_0}{2n}\right)^{n}
 U^\dagger_4
 \left(1\!-\!\frac{aH_0}{2n}\right)^{n}\left(1\!-\!a\delta H\right) G_t
 \quad (t>1) .
\label{tevolve}
\end{eqnarray}
$H_0$ is the kinetic energy operator, the lowest order (in $v^2/c^2$)
term in the Hamiltonian:
 \begin{equation}
 H_0 = - {\Delta^{(2)}\over2\mbox{$M_b^0$}}.
 \end{equation}
The correction terms to the Hamiltonian that we include in
$\delta H$ are $\cal{O}$$(v^4/c^4)$.
They comprise relativistic corrections to the spin-independent $H_0$ as
well as the first spin-dependent terms that give rise to spin-splittings
in the spectrum.
 \begin{eqnarray}
\delta H
&=& - c_1 \frac{(\Delta^{(2)})^2}{8(\mbox{$M_b^0$})^3}
            + c_2 \frac{ig}{8(\mbox{$M_b^0$})^2}\left({\bf {\bf \Delta}}\cdot{\bf E} -
{\bf E}\cdot{\bf {\bf \Delta}}\right) \nonumber \\
 & & - c_3 \frac{g}{8(\mbox{$M_b^0$})^2} \mbox{\boldmath$\sigma$}\cdot({\bf {\bf \Delta}}\times{\bf E} -
{\bf E}\times{\bf {\bf \Delta}})
 - c_4 \frac{g}{2\mbox{$M_b^0$}}\,\mbox{\boldmath$\sigma$}\cdot{\bf B}  \nonumber \\
 & &  + c_5 \frac{a^2{\Delta^{(4)}}}{24\mbox{$M_b^0$}}
     -  c_6 \frac{a(\Delta^{(2)})^2}{16n(\mbox{$M_b^0$})^2} .
\label{deltah}
\end{eqnarray}
The last two terms in $\delta H$ come from finite lattice spacing
corrections to the lattice laplacian and the lattice time derivative
respectively  \cite{gpl2}.  ${\bf {\bf \Delta}}$ is the symmetric lattice derivative and
 ${\Delta^{(4)}}$ is a lattice version of the continuum
operator $\sum D_i^4$. We used the standard traceless cloverleaf operators for
the
chromo-electric and magnetic fields, $g{\bf E}$ and~$g{\bf B}$.  The
parameter~$n$
is introduced to remove instabilities in the heavy quark
propagator caused by the highest momentum modes of the theory.

We tadpole-improve  \cite{tad} our lattice
action by dividing all the gauge fields, $U$, that appear in ${\bf E}$, ${\bf B}$, and
 the covariant lattice derivatives
fields by $u_{0P}$, the fourth root of the plaquette.
This is most easily done as the $U_\mu$'s
are read by the code that evolves propagators.
Tadpole-improvement of the action allows us to
work with tree-level values for the $c_i$'s in $\delta H$ (i.e. 1)
without, we believe,
having to worry about large renormalizations  \cite{colin}. Hence our lattice
action
depends only on two parameters, the bare mass~$\mbox{$M_b^0$}$ and the QCD coupling
constant,~$g$.

Table \ref{params} shows the parameters used in the calculations at 3 different
values of $\beta$.
The configurations were all generated using the standard unimproved Wilson
plaquette action and generously made available to us by the UKQCD
collaboration \cite{ukqcd} and
by Kogut {\it et al} \cite{Kogut}. The results described here at
$\beta$ = 6.0 agree with our previous results \cite{oldups} but
generally have
higher precision, because of an increased number of sources on different
time slices and the increased length of the lattice in the time direction.

Once the quark propagators have been calculated
it is straightforward to obtain anti-quark
propagators and meson correlation functions.
We used the standard interpolating operators described in \cite{oldups} with
source and sink `smearing functions'. We worked in Coulomb gauge and took
wavefunctions for
smearing functions, either from a Richardson potential
($\beta$ = 6.0 and 6.2) or from a Coulomb potential (with modifications,
$\beta$
= 5.7). We took a ground state wavefunction and 2(1) radial excitations
for S states at $\beta$ = 6.0 and 6.2 ($\beta$ = 5.7). For P states we
used a ground state wavefunction and 1(0) radial excitations. We also used
local sources and sinks which were delta functions for S states and
combinations of delta functions for higher orbital excitations.
In addition we looked at S-state mesons with
small non-zero momenta. In the following discussion correlation functions
at zero momentum will be denoted $(^{2S+1}L_J)_{ab}$ where $a$ is the source
smearing function and $b$ the sink smearing function with $l$ for
a delta function (local operator), 1 for the ground state, 2 for the first
excited state
and so on.

At $\beta$ = 5.7 we summed over both initial quark spins.
At $\beta$ = 6.0 and 6.2 we
saved CPU time by fixing the initial quark spin to $+ 1/2$, since the spin-flip
operators in the Hamiltonian are suppressed with respect to $H_0$.
We then used the strong correlations between different polarisations to
obtain reduced errors on the spin splittings for P states \cite{oldups}.

We used 8 different spatial origins for our quark propagators at (2)(4)(1)
different time slices at $\beta$ = (5.7)(6.0)(6.2) to improve
statistics. At $\beta$ = 6.2 all 8 spatial origins were handled simultaneously
by using Z(2) noise at each origin, 1 set per configuration \cite{rdk}.

As described in \cite{oldups} we used multi-exponential fits to the
multiple correlation functions obtained by different combinations of
source and sink. This allowed us to obtain
ground state energies and one or two excited state energies. Two different
types of fit were employed; the `matrix' fit and the `row' fit. The matrix fit
used the matrix of correlators obtained with ground and excited state
sources and sinks. The row fit used the row of correlators with ground and
excited state sources and local operator sinks. We found the correlators
with both local source and sink to be of very little use in fitting.
We had a large number of measurements in every case and so did not run
into problems with our covariance matrix, even for multi-exponential fits
with several parameters. The two different fits gave consistent
results within the errors that we quote.

Final fitted values were chosen by monitoring the quality of the fits (Q) for
given ranges of fitting time, $t_{min}$ to $t_{max}$, as well as
the stability of the fitted parameters. For a given type
of fit, Q generally increases sharply with $t_{min}$ until it reaches a
plateau.
The first fit for which this happens is taken as the preferred value.
In general the $n$th excited energy is taken as reliable from a fit
to $n+1$ exponentials. Table \ref{fits} shows the quality of our fitted results
for a 2 exponential fit to 2 correlators for the $^3S_1$;
$(^3S_1)_{1l}$ and $(^3S_1)_{2l}$, at each $\beta$ value.
Notice how Q increases
from small values of $t_{min}$ as contamination from a third state dies
away. Notice also how stable the fitted ground state energies are for a
very large range of $t_{min}$ values. 

It is interesting to study how the noise in the meson correlators changes
with $\beta$. We expect the ground state meson correlators ($^1S_0$) to
have noise governed by the same mass as the signal since it is the
lightest mass available  \cite{noise}. This means that the errors
in an effective mass
plot will not grow with lattice time. Figure 1 shows this clearly
for the effective masses from the $(^1S_0)_{1l}$ correlator.
The size of the errors at different values of $\beta$ reflects partly
the different statistics available for the different sets of configurations
(see Table \ref{params}). If we multiply the errors at $\beta$ of 5.7 by
$\sqrt{2}$ and at 6.0 by 2, for the different number of
time origins (assuming these are independent), then the errors
are in the ordering $5.7 > 6.0 > 6.2$. The error depends on the
overlap between the squared correlation function and two
$^1S_0$ particles \cite{noise}.
On coarser lattices, the local sink will provide a better
overlap with two $^1S_0$ particles than on finer lattices and so
we would expect the error to be larger. If instead we compare
correlation functions in which the ground state smearing is applied
at both source and sink, $(^1S_0)_{11}$, then the errors at all three values of
$\beta$ are very similar when adjustments for statistics are made
as above. This reflects the fact that the noise should not change
if the physical overlap with two $^1S_0$ states doesn't change.

For higher states than the ground state the noise grows exponentially
with time according to the splitting between that state and the
$^1S_0$. Figure \ref{pmeff} shows this effect for the $^1P_1$ correlator with
ground state smearing at the source and a local sink.
We expect the doubling time for the error to be $\ln(2)/(1P-1S)$ =
1.6 ${\rm GeV}^{-1}$, and this is roughly true at all three values
of $\beta$.
Again the absolute size of the error at fixed physical time is
very similar between all 3 $\beta$ values when adjustments for the
different statistics
are made as above. However, on the coarser lattice many fewer
lattice time points occur before the noise grows overwhelmingly large.

Table \ref{table_energies} shows energies in dimensionless
units obtained from our fits at each value
of $\beta$. 3 exponential fits were used in general so the
values for the $3S$ states should be used with some caution.
$^3 \overline{P}$ is the spin average of the $^3P_{0,1,2}$
states defined by
\begin{equation}
^3 \overline{P} = \frac { 5 M(^3P_2) + 3 M(^3P_1) + M(^3P_0) } {9}.
\end{equation}
This is obtained by measuring spin splittings (see below)
between the $^3P$ states and
the $^1P_1$, and the energy of the $^1P_1$.

Because the quark mass term is missing from our Hamiltonian the
zero of energy becomes shifted so that the energies measured in the
simulation and given in Table \ref{table_energies} cannot be
directly converted to hadron masses.
Differences in energy can be converted directly to physical units using
a value for the lattice spacing, but to obtain absolute masses we
need to know the energy shift. It is sufficient to calculate an absolute
mass for one meson only, and the one for which the most accurate calculation
can be done and compared to experiment is the $1{^3S_1}$, the $\Upsilon$.

To calculate the absolute mass of the $\Upsilon$ we measure the dispersion
relation from the energy of meson correlation functions at small, non-zero
momenta, and fit to a non-relativistic energy-momentum form:
\begin{equation}
aE_{\Upsilon}(p) = aE_{NR, \Upsilon} + \frac {a^2p^2} {2aM_{\Upsilon}} - C_1
\frac {a^4p^4} {8 a^3 M_{\Upsilon}^3}.
\end{equation}
$E_{NR}$ is the energy at zero momentum normally measured (and
given in Table \ref{table_energies}). $C_1$ is a constant to be 
obtained from the fit. $aE(p)-aE_{NR}$ is obtained accurately by
a single exponential fit to the
bootstrapped ratio of correlators at finite and zero momentum. We
use the lowest 1 or 2 non-zero momenta in the fit.
Table \ref{table_energies} shows the kinetic
masses in lattice units, $aM_{\Upsilon}$, obtained at the 3 different
values for $\beta$ for the bare quark
masses given in Table \ref{params}. The value at $\beta$ = 6.0 is taken from
Ref. \cite{oldups} and has not been recalculated
on the Kogut {\it et al} configurations.

P-wave spin splittings can also be obtained most accurately from
ratio fits. Single exponential fits are performed to the ratio of
appropriate polarisation components to maximise the
correlations, as discussed above. Table \ref{table_energies} shows the results
for different splittings in lattice units at the 3 different
values of $\beta$.  Table \ref{psplits} gives a more detailed breakdown for
different polarisation components of the $1P$ fine
structure at $\beta$ = 6.0 and 6.2. For
a given splitting there is no significant difference between
different polarisation components so we average to get a final value
and allow for variations in the error. We see no significant difference
between $^3P_2T$ states and $^3P_2E$ states. This has been checked
explicitly by taking ratios of those correlators.
For the hyperfine splitting ($\Upsilon - \eta_b$)
we are able to
extract both ground and excited splittings (for the first time
in a lattice calculation) from simultaneous fits to the $3 \times 3$
matrix of correlators for the $^1S_0$ and $^3S_1$. These results are also
given in Table \ref{table_energies}.

The wavefunction at the origin is calculated from the ratio
of amplitudes
of row and matrix fits as described in ref. \cite{oldups}.

\section{Discussion}

\subsection{Setting the scale}

One of the useful features of the spectrum of heavy quark bound states
is that the splittings between radial and orbital excitations, spin-averaged,
are to a good approximation independent of quark mass in the
region between bottom and charm. Since not all the bottomonium fine structure
has been seen experimentally, this statement relies to some extent
on estimates of the spin splittings that have not been measured.
However, since spin splittings are very small for bottomonium systems
($\approx 10\%$ of spin-averaged splittings), we still expect little
quark mass dependence for radial and orbital splittings in the region
of $M_b$ when non-spin-averaged splittings are used.
This allows us
to set the scale from lattice calculations, independently of the
requirement to tune the bare lattice quark mass to get the right kinetic
mass for the $\Upsilon$.

In Table \ref{scales} we show values
for the lattice spacing, obtained by fixing various
radial and orbital splittings
to experiment (spin-averaging where possible),
at the 3 different values of $\beta$. These lattice spacing determinations are
very accurate ones for setting the scale in the determination of
$\alpha_s$ \cite{alpha}. Notice that different splittings at a given value of
$\beta$
give slightly different values for $a^{-1}$. This is a feature of the quenched
approximation
which we return to below. First we describe how the determination
of $a^{-1}$ is done and how errors are assigned.

Since the NRQCD action of eq.~\ref{deltah} is corrected
through $\cal{O}$$(a^2)$ for discretisation errors
we would also like to remove other $\cal{O}$$(a^2)$
errors that come from using gluon fields generated with the
simple plaquette action. Fortunately, these
errors can be corrected for
after the calculation.
Perturbatively the correction appears as a shift to
energies and is related to the
wavefunction at the origin. It can then be written in terms of
the hyperfine splitting for $s$ states (for $p$ states
the shift is zero). We use  \cite{alpha}
\begin{equation}
a\,\Delta M_g = \frac{3}{40}\,\left(a\,M_b\right)^2\,a\,\Delta M_{\rm
hyp},
\end{equation}
with $M_b$ set to 5 GeV. The resulting shift to the
splittings of Table~\ref{scales} is given in column 4. This
shift is added to the splittings in lattice units before they
are divided into the physical splitting to obtain $a^{-1}$.
For the shift for 2S states we use the ratio of 2S to 1S hyperfines
determined at $\beta$ = 6.0 and given in Table~\ref{table_energies}.
The statistical error in the splitting is then inflated by
$a\Delta M_g/2$ before calculating the statistical error in
$a^{-1}$ given in column 5.

This determination of $a^{-1}$ also has systematic errors, as
in all lattice determinations. We attempt here to quantify the
errors relevant to our calculation.
There are two sources, physical and unphysical.
The errors from higher order relativistic corrections which have
been ignored are physical and will give the same percentage error
at all values of the lattice spacing.
The unphysical systematic errors come from higher order
discretisation corrections that have not been included. These are
all much larger at the coarsest lattice spacing than
elsewhere.

As a guide to estimating the size of these
errors we have estimated, using a potential model \cite{gpllat91}, the size
of the shifts in energy
caused by the relativistic and discretisation corrections of
$\delta H$ that we {\it have} included and compared those to the
results of NRQCD calculations \cite{oldups}.
The potential model estimates for the sum of relativistic
and discretisation corrections in $\delta H$ at $\beta$ = 6.0
give a
resulting shift to the $b\overline{b}$ $1P-1S$ splitting of $-10$ MeV, made up
of
$-10$ MeV from the relativistic corrections (10 MeV from
the $p^4$ term and $-20$ MeV from the $\vec{D}\cdot\vec{E}$ term)
and cancelling contributions
each of around 10 MeV in magnitude from the two discretisation
corrections \cite{gpllat91}. The $2S-1S$ splitting has smaller
shifts because the expectation values on which the shifts
depend are more similar for the $2S$ and $1S$ than the $1P$ and $1S$.
The expected result from adding all the terms is still $-10$ MeV.
The NRQCD results show
a $1P-1S$ splitting that is 20(30) MeV larger without
$\delta H$ than with, and a $2S-1S$ splitting that is
5(25) MeV larger. This is in good agreement with expectations, albeit with
statistical errors that are too large to show a clear effect.
However, if the shifts had been much larger than the estimates,
they would have been visible above the noise.

{}From this result we can extrapolate
to the size of relativistic and discretisation corrections that
we have not included. The terms that we can estimate most readily
are those involving powers of quark momenta since these are
easy to relate to lower order terms. The terms involving
chromo-electric and magnetic fields and those terms with a structure
that appears for the first time at higher order are
much harder and we have not estimated these.
There is no reason, however, to suppose from
our study above (which
compares the $p^4$ and Darwin terms), that
these terms should be any larger than the ones we can easily
estimate.

Higher order relativistic
corrections would appear as $v^6$ terms in $\delta H$, i.e.
$\cal{O}$$(v^4)$ relative to the leading terms. The percentage
error we expect is then na\"{\i}vely $10\%^2 = 1\%$.
On the other hand, the estimates above using potential
models \cite{gpllat91}, of the
$v^4$ spin-independent
relativistic corrections that we {\it have} included show them each to
be less than half of the 10\% (= 50 MeV) na\"{\i}vely expected.
In addition we actually need the difference between the corrections
for, say $1P$ and $1S$, to get the shift in the splitting.
This indicates that higher
order corrections could be smaller than 1\% too.
Another type of similar higher order correction
is that from radiative corrections to the $c_i$ coefficients
beyond tadpole-improvement.
These should appear
at the level of $\alpha_s$ at an ultra-violet scale
times the $v^4$ relativistic
corrections, giving 0.5\% at $\beta$ = 5.7 and less at higher $\beta$
values.
To encompass both these higher order physical corrections a
1\% error is given as the second error in column 5 of Table~\ref{scales}.

The NRQCD action, eq.~\ref{deltah}, includes the
leading $a$ and $a^2$ corrections which appear in power
counting form
as $p^2a^2v^2$ and $Kav^2$  relative to the leading order
$v^2$ term ($H_0$). $p$ and $K$ are a typical momentum and
kinetic energy associated with the bound state.
Estimates of these terms using potential models \cite{gpllat91} yield
shifts of 40 MeV at $\beta$ = 5.7, 10 MeV at $\beta$ = 6.0
and 5 MeV at $\beta$ = 6.2, in the $1P-1S$ splitting.
Higher order discretisation errors not included could be
radiative corrections to those included (beyond the
tadpole-improvement of these terms which has been done) and we can
estimate these as $\alpha_s(\pi/a)$ \cite{alpha} times the leading errors.
This gives 7 MeV at $\beta$ = 5.7, 1.5 MeV at $\beta$ = 6.0 and
0.7 MeV at $\beta$ = 6.2. Higher order terms in $a$ (such as
$p^4a^4v^2$ terms) would give a percentage
effect roughly the square of the leading order terms, i.e
4 MeV at $\beta$ = 5.7, 0.2 MeV at $\beta$ = 6.0 and
essentially 0 at $\beta$ = 6.2. We should also consider
the first discretisation corrections to the first
relativistic corrections i.e. terms of order $p^2a^2v^4$.
These are very similar looking terms to the $p^4a^4v^2$ terms
and so we can use this to estimate their size. Including powers
of $Ma$ and numerical factors we get a similar size correction
at $\beta$ = 5.7. At higher values of $\beta$ these corrections
are actually more important than those which are higher order
in $a$ but not suppressed by powers of $v^2$. However, all
discretisation corrections become smaller at higher $\beta$
so they are still negligible; 0.5 MeV at $\beta$ = 6.0 and
$0$ at $\beta$ = 6.2.
The third error given in column 5
of Table~\ref{scales} is then conservatively estimated by the sum
of the three errors given above from discretisation errors
in the NRQCD action.
It is interesting to note that
the discretisation corrections are smaller than the anticipated higher
order relativistic corrections except at $\beta$ = 5.7. Note also
that the statistical errors are generally larger than the systematic
- to see any effect from including higher order terms we
would have to improve our statistical error significantly.
Higher order discretisation corrections from the gluon action are anticipated
to be negligible given the size of the $\cal{O}$$(a^2)$ correction
in Table~\ref{scales}.

The $a^{-1}$ determination at $\beta$ = 6.0
agrees with our previous determination \cite{oldups} and at
$\beta$ = 6.2 agrees with previous UKQCD results \cite{cambridge}.

\subsection{Determining the quark mass}

As discussed above, experimental spin-averaged radial and orbital splittings
are very insensitive to the value of the quark mass.
However, in the quenched approximation
there is some mass dependence for these splittings on
the lattice \cite{bc}, and this
is increased if spin-averaging is not done. It is therefore true that a tuned
bare quark mass is necessary to get the right radial and orbital
splittings. The spin splittings are much more sensitive to the
quark mass (roughly as its inverse), and it is essential
to tune the quark mass to get these correct.
The way in which we tune the quark mass is to adjust it until
the kinetic mass of the $\Upsilon$ agrees with experiment.
Table~\ref{kmass} shows the kinetic mass of the $\Upsilon$ (given
in lattice units in Table~\ref{table_energies}) in GeV, for
each of the 3 values of $\beta$, using $a^{-1}$ from Table~\ref{scales}. There
is
reasonable agreement with the experimental result
9.46 GeV in each case, provided that the value for
$a^{-1}$ is taken from the $\Upsilon^{'} - \Upsilon$ splitting.
The systematic errors in the determination
of $aM_{kin}$ are at the 1\% level from the same sources as systematic
errors in the determination of $a^{-1}$. This is smaller in every case than
the statistical error, dominated by the statistical
uncertainty in $a^{-1}$.

The difference between the kinetic mass of a meson and its energy at zero
momentum is calculable in perturbation
theory \cite{colin_again}. The formula which relates $E_{NR}$ and
 the kinetic mass
$M$ is :
\begin{equation}
M = 2(Z_m M^0_b - E_0) + E_{NR}
\label{shift}
\end{equation}
where $Z_m$ is the mass renormalisation and $E_0$ the energy shift.
Table~\ref{bmass} gives values for $Z_m$ and $E_0$ appropriate to the
different bare masses used at each value of $\beta$. Tadpole-improved
lattice perturbation theory has been used for these parameters and the scale
of $\alpha_s$ set using the BLM scheme \cite{colin_again}. The values obtained
for $M_{\Upsilon}$ from the perturbative expression, (\ref{shift}),
are given in lattice units in the sixth column
and should be compared with the results for $aM_{kin}$ in Table~\ref{kmass}.
There should be
agreement at all values of $\beta$ independent of whether the quark mass
is well tuned to that appropriate to the $b$ or not, and we
see that there is. The perturbative error is at
$\cal{O}$$(\alpha_s^2)$ and is taken here as the square
of the $\cal{O}$$(\alpha_s)$ term. Note that the relationship
in eq.~\ref{shift} is well defined perturbatively.
Non-perturbative values for the shift between $E_{NR}$
and $M$ can also be measured on the lattice for, say,
$\Upsilon$ and used, divided by 2, for $B$ physics \cite{us_fb}.

We can use the information above to determine the
$b$ quark mass \cite{mb} in two independent ways and
ask whether the renormalised $b$ quark mass
we obtain
scales from one value of $\beta$ to the next.
These results will be presented elsewhere \cite{mass_in_prep}.

\subsection{Radial and Orbital splittings}

{}From the results of Table~\ref{table_energies} we can
investigate the scaling of
spin-averaged splittings across the
range of $\beta$ values we have
used. Our analysis of systematic errors in section 3.1
already implied that we do not expect to see
violations of scaling in these quantities.

Figure~\ref{upsrho} shows with circles the ratio
of our $\overline{\chi_b} - \Upsilon$ splitting
to the $\rho$ mass, obtained by the UKQCD collaboration
at the same three values of $\beta$ \cite{ukqcd,rowland}.
The $\rho$ mass was calculated using an action with $\cal{O}$$(a)$
errors removed using a clover term with tadpole-improved coefficient;
i.e. a similar philosophy to that used here to remove
discretisation errors from the $\Upsilon$ splittings. The remaining errors
in the $\rho$ determination may then be of $\cal{O}$$(a^2)$.
For the $1P -1S$ splitting
we have removed the $\cal{O}$$(a^2)$ gluonic errors so,
as discussed earlier, the remaining errors are higher order
in $a$ and/or suppressed by powers of $v^2$.
The plot is very flat showing that good scaling is
obtained. We can rule out scaling violations in this ratio with a scale
larger than a few hundred MeV, and it would be very unlikely for
the $\rho$ mass and the $\Upsilon$ system to have identical
and cancelling large scaling violations.
The plot also shows in the comparison
to the GF11 results, that if an
unimproved calculation is done of the $\rho$ mass \cite{gf11}, an
absence of scaling is quite evident (in this case linear
in $a$).

The ratio $\Delta (\overline{\chi_b} - \Upsilon) / m_{\rho}$
shows a big discrepancy
with experiment. This we believe is an
error from the quenched approximation. The
scales intrinsic to a light hadron system and the
$\Upsilon$ are quite different and the
coupling constant does not run correctly between these scales
in the quenched approximation.
Again, these errors are masked by discretisation errors
for an unimproved light hadron spectrum.

Figure~\ref{upsrho} also shows the ratio of the
$\overline{\chi_b} - \Upsilon$ splitting
to the $\Lambda$ parameter of QCD in the V scheme \cite{tad}.
Again good scaling is seen over this limited range of lattice
spacings.

Figure~\ref{siratios} shows the dimensionless ratios of various
splittings within the $\Upsilon$ system (spin-averaged where
possible) as a function of lattice spacing.
The ratios shown are for the $\Upsilon^{''} - \Upsilon$,
$\Upsilon{'} - \Upsilon$ and the $h_b^{'} - \Upsilon$ (experiment for the
latter case using $\overline{\chi_b^{'}}$ for $h_b^{'}$).
The ratios
are constant as a function of lattice spacing, although the
statistical errors are rather large for the 3S and 2P cases.
This means that we can
interpret our results as continuum results.
Note that our current value for the energy of the
$2^3S_1$ at $\beta$ = 6.0 has dropped by $1 \sigma$ from
our previous calculation \cite{oldups} with slightly smaller
statistical errors. Our $1^1P_1$ energy has not
changed and the error has fallen by a factor of 2.
Our best value for the ratio $(2S-1S)/(1P-1S)$
is now 1.35(6) and the discrepancy with experiment
remains $(1 - 2) \sigma$, controlled by the
uncertainty in the $2^3S_1$ energy. The discrepancy with experiment
for the ratios involving the 3S and 2P
could be affected by finite volume errors as well
as quenching errors.

\subsection{Spin splittings}

Spin splittings in lattice units are given in
Table~\ref{table_energies}.
This includes, for the first time, the radially excited hyperfine
splittings, $\Upsilon^{'} - \eta_b^{'}$ and $\Upsilon^{''} - \eta_b^{''}$
at $\beta$ = 6.0 where our results are most precise. No
hyperfine splitting has yet been seen experimentally for
the $\Upsilon$ system, and a radially excited hyperfine
splitting is still awaited for charmonioum. We find the 2S
hyperfine splitting to be about half that of the ground state
here. A proper analysis however, has to study the lattice spacing
dependence and the effect of unquenching on this ratio.

In this section we investigate the dependence in physical units of
the ground state spin splittings on the
lattice spacing. We also compare to recent UKQCD results \cite{manke}
in which a systematic study at next order in relativistic
and discretisation corrections has been done for
spin-dependent terms, i.e. those which affect the spin splittings,
only.
By comparing the scaling of our results with their results we
can untangle to some extent the difference between the discretisation
errors at fixed $a$, which are unphysical, and the relativistic
corrections, which are physical. Whether or not the spin-dependent
relativistic corrections (i.e. terms of $\cal{O}$$(Mv^6)$) are
sizeable or not is important since these are terms outside the scope
of potential models.

To investigate scaling
we show in Figures~\ref{hyprat} and~\ref{prat}
various spin splittings in MeV setting
the scale from the
$\Upsilon^{'} - \Upsilon$ splitting as a function
of $a^2$. The $\Upsilon^{'} - \Upsilon$ splitting is used since
this is the one for which our quark masses are closest to being
tuned and this is important for spin splittings \cite{a_note}.
The results from Table~\ref{table_energies} are given as squares.
We include in the Figures the direct uncertainty in the spin
splittings from statistical errors in the lattice spacing.
There is an additional error, which we have not included, from
the uncertainty in the kinetic mass from the error in the
lattice spacing. This uncertainty leads to an uncertainty
in the tuned quark mass which affects the spin splittings.
If the spin splittings vary as the inverse quark mass (roughly
true at least for the hyperfine splitting), then
this second uncertainty is actually equal in size, correlated
with and in the
same direction as the error that we have included. This
would lead
to an approximate doubling of the error bars in the Figures.

Clear dependence on the lattice
spacing is visible for the hyperfine splitting,
$\Upsilon - \eta_b$.
For the $1P$ fine structure
it is less clear because statistical errors, particularly
at $\beta$ = 6.2, are so much higher.
Lattice spacing dependence in our results is not surprising
since the Hamiltonian we have
used has only leading order spin-dependent
terms with no corrections for
discretisation errors in these terms. These errors are expected to
be $\cal{O}$$(a^2)$ relative to the leading term
and therefore in relative terms significantly larger than
for our spin-independent splittings. In addition, in a potential
model picture the fine structure
is provided by potentials which are generally of much shorter range
than the central potential so we would expect the fine structure to
have a harder scale for discretisation errors
than the radial and orbital
splittings.
If we parameterize the scaling
violations in the hyperfine splitting by  \cite{john}:
\begin{equation}
\frac {splitting} {2S-1S} = \left(\frac {splitting} {2S-1S}\right)_{0} ( 1 -
\mu^2 a^2), \label{scaling}
\end{equation}
we find the parameter $\mu$ to take a
value around 0.8 GeV for the squares.

For the $1P$ fine structure we have the advantage that
we can compare to experimental results as well
as comparing results at different values of the
lattice spacing. In Figure~\ref{prat}
we notice a clear disagreement with experiment for the
squares in that both
the overall scale of splittings is too small and the
ratio
\begin{equation}
\rho = \frac {\Delta(^3P_2 - ^3P_1)} {\Delta(^3P_1 - ^3P_0)}
\label{rho}
\end{equation}
is too large. Similar results were noticed in our
charmonium spectrum calculation \cite{charm}. The
disagreement with experiment over $\rho$ was suggested there
as a discretisation effect and results here tend to confirm
that discretisation errors do have an effect in this ratio.
Our value for $\rho$ at beta=5.7 is 1.4(3) and at
6.0, 1.1(3). The experimental value is 0.66.
The comparison between the squares at $\beta$ = 5.7 and 6.0
shows that scaling violations are just about visible
for the $^3P_0$ and $^3P_2$ above the statistical errors.
The slope of the scaling violations is similar to that
for the hyperfine splitting (but with large errors).
Also in agreement with the hyperfine splitting we can see that
adding discretisation corrections would increase the splittings
and this would make the 1P fine structure closer to experiment.

On Figures~\ref{hyprat} and ~\ref{prat} we
also show results from Manke {\it et al} \cite{manke}
at $\beta$  = 6.0 and 5.7 with diamonds.
They have used the same bare quark masses as us
at both values of $\beta$ but included higher order
spin-dependent relativistic ($\cal{O}$$(v^6)$) and
discretisation ($\cal{O}$$(a^2v^4)$) corrections to the
action \cite{gpl2}, all tadpole-improved with
$u_{0P}$. We take their spin splittings in
lattice units and convert to physical units using our
$\Upsilon^{'} - \Upsilon$ splitting since, as explained earlier,
this is the $a^{-1}$ for which the quark masses are best
tuned.
Our result for this splitting should be the same as theirs
since they have added only spin-dependent terms \cite{v6_note} and
ours is more precise.
The lattice spacing dependence of their results
should be much reduced over ours, and we see that
it is. By comparing their results and ours,
we can see that the relativistic corrections
act in the opposite direction
to the discretisation corrections, in agreement with the
findings of \cite{trottier}.

By comparing the results of \cite{manke} with ours
we can attempt to untangle the size of $v^6$ relativistic
corrections to the fine structure.
First, however, we should also include radiative corrections to
the leading order terms, beyond tadpole improvement.
These are of $\cal{O}$$(\alpha_s v^4)$
and therefore of a similar size to the relativistic
$v^6$ corrections.
The $c_i$ in equation 3 take the form
$1 + c_i^{(1)}\alpha_s(q^{*}) + \ldots$ with $c_i^{(1)}$ of
$\cal{O}$$(1)$.
Preliminary perturbative results
are available \cite{trottier2} for $c_4^{(1)}$,
for the $\vec{\sigma}\cdot\vec{B}$ term in
the action, and confirm this
expectation. In perturbation
theory the hyperfine splitting then picks up a factor
of the square of this correction, so it is quite sensitive to
$c_4^{(1)}$.
Figure~\ref{hyprat} shows with fancy squares our results
for the hyperfine splitting rescaled by
$(1+\alpha_P(3.4/a)c_4^{(1)})^2$. Values for $\alpha_P$ were
taken from \cite{alpha}.
$c_4^{(1)}$ increases as $Ma$ increases, which
gives a larger correction at low $\beta$. Since this is also
the direction in which $\alpha_P(3.4/a)$ increases the correction
at $\beta$ = 5.7 is large, 50\%. At $\beta$ = 6.2 the correction
from  $c_4^{(1)}$ is 20\% \cite{c4_note}.
The fancy diamonds show the rescaled results from \cite{manke}
in which we have simply rescaled the leading order piece
i.e. we have applied the same shift in MeV to these results
as to ours. It is clear from this Figure that the
radiative corrections act in the same direction as
discretisation corrections discussed earlier and absorb some of them,
softening the value of $\mu$ in equation~\ref{scaling}.

For the hyperfine splitting, we will assume that the
discretisation corrections at $\beta$ = 6.0 in the
$(v^6, a^2v^4)$ action \cite{manke} are of $\cal{O}$$(a^4)$ and
therefore negligible already at $\beta$ = 6.0.
For the $(v^4)$ action (this work) we will take the result at
$\beta$ = 6.2 as being the least affected by discretisation
errors (we don't expect more than a 5\% effect here).
The difference then between the result in physical
units between $\beta$ = 6.2 ($v^4$) and $\beta$ = 6.0
($v^6,a^2v^4$) represents the effect of relativistic
corrections. From Figure~\ref{hyprat}
with fancy squares and diamonds
this gives a 15\% decrease in the hyperfine
splitting and from the plain squares and diamonds a 20\%
decrease, consistent with the expectation
of around 10\% from power counting in $v^2/c^2$.

For the $p$ fine structure, the radiative corrections
are not available and they would in any case be much
harder to apply without doing the full lattice
calculation including them.
We therefore just compare the results with (diamonds) and
without (squares) the $v^6$ and $a^2v^4$ corrections in Figure~\ref{prat}.
The lattice spacing
dependence is apparently reduced by the
$a^2v^4$ corrections, although the statistical errors
make it hard to quantify this result. The relativistic
corrections again act to reduce the splittings, now
to something like 50\% of experiment. The size of the relativistic
corrections is significant, possibly several times
the na\"{\i}ve estimate of 10\%. They could be
offset by radiative corrections, however, in a
fully consistent calculation that includes the next order
in $v^2$, $a^2$ and $\alpha$ over our calculation.
Radiative corrections which increase the spin splittings,
as for the hyperfine case, would improve the
agreement with experiment, but unless these are
surprisingly large,
the splittings will remain too small in the quenched
approximation.
Initial unquenched results for the $v^4$ action \cite{alpha}
and the $v^6, a^2v^4$ action \cite{SESAM} have indicated
an increase in the $p$ spin splittings,
but currently statistical errors are too large
to clearly determine the size of unquenching effects.

Ref. \cite{manke} points out that
the $(v^6, a^2v^4)$ action has a much improved value
of $\rho$ (eq.~\ref{rho})
over the $(v^4)$ action. They quote 0.56(19)
at $\beta$ = 6.0 in the quenched approximation, now
in agreement with experiment, confirming the
discretisation effects for this ratio discussed above.
In fact, we do not expect this
ratio to agree exactly with experiment in the quenched approximation.
An analysis of $\rho$ for the simple Cornell potential \cite{schladming}
shows that for a quenched $\alpha_s$ which is too small, the
value of $\rho$ is reduced.
With better statistics it may become
clear that the quenched value for $\rho$ is actually smaller than
the experimental result.

Another way to study the sensitivity of the fine structure
to higher order terms in the action
is to change the value of the tadpole coefficient, $u_0$.
Several calculations \cite{improve} have indicated that
the average link in Landau gauge, $u_{0L}$ is to be
preferred over the fourth root of the plaquette, $u_{0P}$,
that we have used here. Changing $u_0$ is equivalent to
resumming various terms to all orders in the $c_i$.
It is again easy to rescale the hyperfine splitting
after the simulation to account for a different value of
$u_0$, by counting the powers of $u_0$ that
appear in front of the $\vec{\sigma}\cdot\vec{B}$ term.
The hyperfine splitting is very sensitive to $u_0$,
scaling as
$1/u_0^6$, when allowing for $u_0$ factors in $H_0$ which change
the renormalised quark mass.
The expected scaling is confirmed by calculations at different
values of $u_0$ \cite{oldups} (although here a fixed bare
quark mass was used).
The hyperfine splitting increases most on using $u_{0L}$
at $\beta$ = 5.7. There the ratio of $u_{0P}/u_{0L}$ is
1.045; at $\beta$ = 6.0 it is 1.020 and at $\beta$ = 6.2,
1.013 \cite{jonivar}.
The hyperfine splitting
with $u_{0L}$ then has somewhat
reduced lattice spacing dependence. Note that $c_4$
will be different for the action with $u_{0L}$, since a
different amount of the radiative correction has been
absorbed by the tadpole factors. $c_4^{(1)}$ in fact is
larger, by 0.24, being the cube of the ratio of
the $u_0$ factors to $\cal{O}$$(\alpha)$.
The size of relativistic and discretisation corrections will
change by differing amounts on changing $u_0$ because they
have different effective powers of $u_0$ in their
coefficients.

{}From the above we can conclude that systematic
errors in the fine structure still exceed 10\%, even
in the quenched approximation.
To improve the accuracy the $v^6$ and $a^2v^4$ action
must be used, with radiative corrections included for
the leading spin-dependent terms.
For the hyperfine splitting the sensitivity to the
$c_i$ is such that the $\cal{O}$$(\alpha^2)$ term also
needs to be known (or at least bounded).
The determination of the $c_i$ is in progress \cite{trottier2}.
More accurate meson kinetic masses and lattice spacing
determinations are also required so that
the quark mass can be tuned more accurately, since the
spin splittings are sensitive to this.

The  wavefunction at the
origin is another short distance quantity but
one which does not vanish as the spin-dependent terms are
switched off. In Figure~\ref{wave} we show results
for $\psi(0)$ for
the $\Upsilon$ and $\Upsilon^{'}$ as a function of $a^2$, again
setting the scale from the $\Upsilon - \Upsilon^{'}$ splitting
since the wave function is sensitive to quark mass.
The lattice spacing dependence does not seem severe but
the
statistical errors are large. They are dominated by those from
$a^{-1}$ which here is raised to the 3/2 power.
The bursts show experimental values based
on the leptonic widths and the na\"{\i}ve
Van Royen-Weisskopf formula,
\begin{equation}
\Gamma_{l^{+}l^{-}} = 16 \pi \alpha^2 e_Q^2 \frac {|\Psi(0)|^2} {M_v^2},
\end{equation}
where $M_v$ is the vector meson mass.
There will be radiative corrections to this formula
of $\cal{O}$$(\alpha_s)$. For potential
model wavefunctions these radiative corrections have
coefficient $16/3\pi$ but this does not have to be
the same for NRQCD. Calculations
of the renormalisation
factors have only been done for lower order actions than
the one used here \cite{beth}.
There will also be terms
of $\cal{O}$$(1/M_b, \alpha_s/M_b)$ from mixing of the vector
current with
higher dimension operators (in a similar way to the
mixing for $f_B$ which has been worked out in some
detail \cite{us_fb}).

A lot of the radiative corrections cancel
in the ratio of leptonic widths
for the $1S$ and $2S$ states. Experimentally this
ratio is 2.5(2). Our result at $\beta$ = 6.0 simply from
the ratio of $|\Psi(0)|^2$ is
1.7(4).  We expect errors in the quenched approximation
which would suppress the 1S wavefunction at the origin
relative to the 2S, so that it is not surprising that
our result is somewhat smaller than experiment.
We cannot yet say whether the reduction observed is
reasonable.

To obtain results for the real world from simulations in
the quenched approximation and at 2 flavours of
dynamical quarks, extrapolations in $N_f$ (= the
number of flavours of dynamical quarks) will be required.
For this it is important to have scaling quantities, as
we have discussed here.
It is also necessary to use quantities which are sensitive to the
same number of dynamical fermions. This means that for a given
splitting we should set the scale using a quantity which
has similar internal momentum scales. For example, for the
fine structure it may not be optimal to use the softer
radial and orbital splittings to set the scale. Typical
momentum scales for radial and orbital splittings are
around 1 GeV so they can be expected to `see' 3 flavours
of dynamical fermions in the real world \cite{alpha}.
The fine structure may instead `see' 4 flavors and this
would imply that an extrapolation in $N_f$ can only
be done in terms of a quantity which also sees
4 flavors.

In Figure~\ref{hypp} we show a plot of the hyperfine
splitting in MeV against the square of the lattice spacing using
the $1^3P_2 - 1^3P_0$ splitting to set the scale.
Statistical errors are now very large, despite
using the most accurate of the $P$ fine structure splittings.
Much higher accuracy
will be
needed to use this ratio for $N_f$ extrapolations.

\section{Conclusions}

We have calculated the spectrum of bottomonium
bound states at 3 different values of the lattice spacing using
NRQCD in the quenched approximation.
 Ratios of spin-independent splittings are constant in
this region, independent of the lattice spacing.
This is a necessary
requirement for the results to make physical sense. There is
no necessity to extrapolate to vanishing lattice
spacing (impossible anyway
for NRQCD). The constancy of the results implies that they can now
be used in extrapolations to real-world $N_f$ values
when
combined with results on dynamical
configurations  \cite{alpha,SESAM}.

For spin-dependent splittings there is some visible
(unphysical) lattice spacing
dependence. This is not surprising since our spin
splittings are only
accurate to leading order, given the terms in
the Hamiltonian that we have
used.
A comparison to a calculation \cite{manke} which
does include higher order
relativistic and discretisation corrections to spin-dependent terms
shows that the relativistic spin-dependent corrections could be
as large as 20\% for the hyperfine splitting where
they can be quantified. They act to reduce
spin-dependent splittings
in all cases, and this increases the disagreement with experiment
for the $P$ fine structure in the quenched approximation.

For extrapolations to real world values of $N_f$ we may need to
employ different techniques for fine structure to that for
radial and orbital splittings since different momentum scales
are involved and this may lead to a different number of
dynamical flavors being `seen'. Statistical errors are
currently too high to use
a $P$ spin splitting
to set the scale in such a programme.

\acknowledgements

This work was supported by PPARC, by the US
Department of Energy (under grants DE-FG02-91ER40690, DEFG03-90ER\-40546,
DE-FG05-84ER40154, DE-FC02-91ER75661) and by the National
Science Foundation. The numerical calculations
were carried out at NERSC,
at the
Rutherford Laboratory Atlas Centre and at the Edinburgh Parallel
Computing Centre under EPSRC grant GR/K55745. We thank UKQCD and Kogut 
{\it et al} for
making their configurations available to us.

We thank Sara Collins, Joachim Hein, Thomas Manke and Howard Trottier for
useful discussions.
CTHD is grateful to the Institute for Theoretical
Physics, UC Santa Barbara for hospitality and to the
Leverhulme Trust and the Fulbright Commission for funding while this work
was written up.


\begin{figure}
\caption[gpg]{Effective mass plots for the $(^1S_0)_{1l}$ correlator
at all three values of $\beta$, in order with
$\beta$ = 5.7 at the top.
The time axis has been converted to physical units of ${\rm GeV}^{-1}$
using the $\overline{\chi_b} - \Upsilon$ splitting to set
the scale (Table \ref{scales}).}
\end{figure}

\begin{figure}
\caption[junk]{Effective mass plots for the $(^1P_1)_{1l}$ correlator
at all three values of $\beta$, in order with
$\beta$ = 5.7 at the top.
The time axis has been converted to physical units of ${\rm GeV}^{-1}$
using the $\overline{\chi_b} - \Upsilon$ splitting to
set the scale (Table \ref{scales}).}
\label{pmeff}
\end{figure}

\begin{figure}
\caption[gjl]{ Dimensionless ratios of the
$\overline{\chi_b} - \Upsilon$ splitting to the parameter
$\Lambda_V$ (diamond) and to the UKQCD
$\rho$ mass \cite{ukqcd} (circle) against the
lattice spacing in fm (set by the $\overline{\chi_b} - \Upsilon$ splitting).
Results using the GF11 $\rho$ mass \cite{gf11} are given by
squares.
The burst represents the experimental value for $\Delta (\overline{\chi_b}
- \Upsilon)/m_{\rho}$.}
\label{upsrho}
\end{figure}

\begin{figure}
\caption[junk]{Dimensionless ratios of various splittings to the
$\overline{\chi_b} - \Upsilon$ splitting against the
lattice spacing in fm (set by the $\overline{\chi_b} - \Upsilon$ splitting).
Circles represent the ratio for the
$\Upsilon^{''} - \Upsilon$ splitting (experiment short
dashes)
and crosses for the $h_b^{'} - \Upsilon$ (experiment using
$\overline{\chi_b^{'}}$ for $h_b^{'}$ dot-dash).
The diamonds show the $\Upsilon^{'} - \Upsilon$ ratio with $a^2$
 gluonic corrections (as described in the text) and the
squares uncorrected results (experiment dashed line). The squares and 
crosses have been offset slightly in the horizontal direction for 
clarity. }
\label{siratios}
\end{figure}

\begin{figure}
\caption[flkml]{ The hyperfine splitting in MeV using the
$\Upsilon^{'} - \Upsilon$ splitting to set the scale, vs $a^2$ in
${\rm fm}^2$.
Plain squares indicate our results from Table~\ref{table_energies}.
The diamonds indicate
the results from ref. \cite{manke} using a higher order
action. They are at matching values of $\beta$ but
offset slightly for
clarity.
Fancy squares indicate our results, rescaled by the square of $c_4$
calculated to $\cal{O}$$(\alpha)$ in \cite{trottier2}.
The fancy diamonds indicate
the result from ref. \cite{manke}
shifted by the same amount as our results
to account for radiative corrections to the $\sigma\cdot B$ term.
The error bars shown are statistical only and include some of the
error from the uncertainty in the scale (see text). The $x$ axis errors
from uncertainty in the scale are shown only for the squares
for clarity. }
\label{hyprat}
\end{figure}

\begin{figure}
\caption[flkml]{ The splitting in MeV between various $1^3P$
states and the $1^3\overline{P}$ using the
$\Upsilon^{'} - \Upsilon$ splitting to set the scale, vs $a^2$ in
${\rm fm}^2$.
Squares indicate our results, diamonds the results of \cite{manke}
at matching values of $\beta$ but
offset slightly for clarity.
The bursts indicate the experimental
values.
The error bars shown are statistical only and include some of the
error from the uncertainty in the scale (see text). The $x$ axis errors
are not shown for clarity - they are of the same size as those in
Figure~\ref{hyprat}.}
\label{prat}
\end{figure}

\begin{figure}
\caption[flkml]{ The wavefunction at the origin in
${\rm GeV}^{3/2}$ vs $a^2$ in ${\rm fm}^2$ setting the scale from
the $\Upsilon^{'} - \Upsilon$ splitting. Points for
the $1^3S_1$ are shown as squares and $2^3S_1$ as
diamonds. The errors shown are statistical only and
the errors on the $x$ axis are shown only for the squares.
The bursts show experimental points derived from
the leptonic widths of the $\Upsilon$ and $\Upsilon^{'}$
using the na\"{\i}ve Van Royen-Weisskopf formula.}
\label{wave}
\end{figure}

\begin{figure}
\caption[flkml]{ The hyperfine splitting in MeV using the
$1^3P_2 - 1^3P_0$ splitting to set the scale, vs $a^2$ in
${\rm fm}^2$. Squares show our results, diamonds the
results of \cite{manke}.}
\label{hypp}
\end{figure}


\begin{table}
\caption[junk]{The parameters used in calculations at 3 different values of
the QCD coupling, $\beta = 6/g^2$.}
\begin{tabular}{cccccccc}
& $aM^0_b$ & $n$ & $u_{0P}$ & $V$ & \# confs. & \# sources & collab.\\
\hline
$\beta$ = 5.7 & 3.15  & 1 & 0.861 & $12^3 \times 24$ &  200 & 8 $\times$ 2 &
UKQCD\\
$\beta$ = 6.0 & 1.71 & 2 & 0.878 & $16^3 \times 32$ & 149 & 8 $\times$ 4 &
Kogut {\it et al}\\
$\beta$ = 6.2 & 1.22 & 3 & 0.885 & $24^3 \times 48$ & 216 & 8 (Z(2)) & UKQCD\\
\end{tabular}
\label{params}
\end{table}

\begin{table}
\caption[junk]{Fitted results for a two exponential fit to two $^3S_1$ correlators,
$(^3S_1)_{1l}$ and $(^3S_1)_{2l}$.
Fitted energies in lattice units
are given with errors as well as the $Q$ value for the fit at all
three values of $\beta$.}
\begin{tabular}{cccccccccc}
$\beta$ & \multicolumn{3}{c}{5.7} & \multicolumn{3}{c}{6.0} &
\multicolumn{3}{c}{6.2}  \\
\hline
$t_{min}$ & $E_1a$ & $E_2a$ & Q & $E_1a$ & $E_2a$ & Q & $E_1a$ & $E_2a$ & Q \\
\hline
2 &&&&&&& 0.3128(3) & 0.517(3) & 0.00 \\
3 & 0.5186(6) & 0.888(5) & 0.01 &&&& 0.3130(3) & 0.508(3) & 0.00 \\
4 & 0.5186(6) & 0.901(7) & 0.37 &&&& 0.3131(3) & 0.504(4) & 0.00 \\
5 & 0.5186(6) & 0.91(1) & 0.64 & 0.4540(2) & 0.717(2) & 0.00 & 0.3131(3) &
0.499(4) & 0.02 \\
6 & 0.5188(6) & 0.93(1) & 0.73 & 0.4539(2) & 0.710(3) & 0.14 & 0.3132(3) &
0.499(5) & 0.04 \\
7 & 0.5188(6) & 0.95(3) & 0.70 & 0.4539(2) & 0.708(3) & 0.14 & 0.3133(3) &
0.493(6) & 0.10 \\
8 & 0.5187(6) & 0.93(4) & 0.65 & 0.4539(2) & 0.705(4) & 0.13 & 0.3133(3) &
0.491(7) & 0.08 \\
9 & 0.5186(6) & 0.93(6) & 0.70 & 0.4539(3) & 0.697(6) & 0.15 & 0.3132(3) &
0.488(8) & 0.07 \\
10 &&& & 0.4539(3) & 0.697(6) & 0.14  & 0.3131(3) & 0.49(1) & 0.07 \\
11 &&& & 0.4539(3) & 0.690(7) & 0.14 & 0.3130(3) & 0.49(1) & 0.06 \\
12 &&& & 0.4538(3) & 0.68(1) & 0.13 & 0.3130(4) & 0.48(1) & 0.05 \\
13 &&& & 0.4537(3) & 0.69(1) & 0.30 & 0.3131(4) & 0.49(2) & 0.05 \\
14 &&& & 0.4537(3) & 0.70(2) & 0.24 & 0.3133(4) & 0.50(2) & 0.04 \\
\end{tabular}
\label{fits}
\end{table}


\begin{table}
\caption[junk]{Fit results for dimensionless $\overline{b}b$ energies
and splittings, $aE$ and $a \delta E$ for the quenched approximation,
at three different values of $\beta$. Below these are given the
kinetic mass (see text) and the wavefunction at the origin, in lattice units.}
\begin{tabular}{llll}
$\beta$ & 5.7 & 6.0 & 6.2 \\
Energies, $aE$ : &&& \\
$1{^1S}_0\ (\eta_b)$ & 0.5029(5) & 0.4415(3) & 0.3028(2)\\
$1{^3S}_1\ (\Upsilon)$ & 0.5186(6) & 0.4537(5)& 0.3132(3) \\
$2{^1S}_0\ (\eta_b^{'})$ & 0.92(3) & 0.678(8) & 0.478(6)\\
$2{^3S}_1\ (\Upsilon^{'})$ & 0.94(4) & 0.686(8)& 0.488(8) \\
$3{^3S}_1\ (\Upsilon^{''})$ & \quad -  & 0.83(3) & 0.65(4) \\
$1{^1P}_1\ (h_b)$ & 0.843(6) & 0.627(3) & 0.438(5) \\
$2{^1P}_1\ (h_b^{'})$ & \quad - & 0.823(14) & 0.60(7) \\
$1{^3 \overline{P}}\ (\overline{\chi_b})$ & 0.845(6) & 0.628(3) & 0.440(5)\\
\hline
Splittings, $a \delta E$ : &&& \\
$1^3S_1 - 1^1S_0$ & 0.01575(8) & 0.01237(14) & 0.01038(14) \\
$2^3S_1 - 2^1S_0$ & \quad - & 0.006(1) & \quad - \\
$3^3S_1 - 3^1S_0$ & \quad - & 0.005(3) & \quad - \\
$1^3P_2 - 1^3P_0$ & 0.020(2) & 0.0147(17) & 0.021(7)\\
$1^3P_2 - 1^3P_1$ & 0.011(2) & 0.0078(15) & 0.010(6)\\
$1^3P_1 - 1^3P_0$ & 0.0079(5) & 0.0069(12) & 0.010(7) \\
$1^1P_1 - 1^3P_1$ & 0.003(2) & 0.0028(6) & 0.003(4) \\
\cline{2-4}
$1^3P_2 - 1^3\overline{P}$ & 0.0059(7) & 0.0042(5) & 0.0058(24)\\
$1^3\overline{P} - 1^3P_1$ & 0.0052(12) & 0.0036(8) & 0.0046(37)\\
$1^3\overline{P} - 1^3P_0$ & 0.0137(11) & 0.0105(10) & 0.0148(52)\\
$1^3\overline{P} - 1^1P_1$ & 0.002(2) & 0.0008(8) & 0.0018(28) \\
\cline{1-4}
Kinetic mass, $aM_{kin}$ : &&& \\
$M(1{^3S}_1)$ & 7.06(7) & 3.94(3) & 2.89(3)  \\
\cline{1-4}
Wavefunction at the origin : &&& \\
$|\psi(0)|a^{3/2}$ for $1^3S_1$ & 0.385(5) & 0.1525(14) & 0.1116(12) \\
$|\psi(0)|a^{3/2}$ for $2^3S_1$ & 0.30(2) & 0.118(14) & \quad - \\
$|\psi(0)|a^{3/2}$ for $3^3S_1$ & \quad - & 0.19(3) & \quad - \\
$|\psi(0)|a^{3/2}$ for $1^1S_0$ & \quad - & 0.1621(13) & 0.1314(12) \\
$|\psi(0)|a^{3/2}$ for $2^1S_0$ & \quad - & 0.115(12) & \quad - \\
$|\psi(0)|a^{3/2}$ for $2^1S_0$ & \quad - & 0.20(3) & \quad - \\
\end{tabular}
\label{table_energies}
\end{table}

\begin{table}
\caption[junk]{Individual $1P$ spin splittings in lattice units
 with  quantum numbers, polarizations and smearing
combinations specified, along with the final value used
in Table \ref{table_energies}. $E$ and $T$ stand for the 
different lattice representations of the continuum spin 2 operator.}
\begin{tabular}{ccccc}
splitting & $\beta$ = 6.0 & result & $\beta$ = 6.2 & result \\
\cline{1-5}
${^3P}_2T_{yz}(1,loc) - {^1P}_1y(1,loc)$ & 0.0060(7) & 0.0050(14) & 0.0072(47)
& 0.0076(42)\\
${^3P}_2T_{yz}(1,1) - {^1P}_1y(1,1)$ & 0.0048(7) &&&\\
${^3P}_2T_{zx}(1,loc) - {^1P}_1x(1,loc)$ & 0.0065(8) & & 0.0060(52) & \\
${^3P}_2T_{zx}(1,1) - {^1P}_1x(1,1)$ & 0.0045(7) &&&\\
${^3P}_2E_{zx}(1,loc) - {^1P}_1z(1,loc)$ & 0.0050(8) && 0.0083(41) &\\
${^3P}_2E_{zx}(1,1) - {^1P}_1z(1,1)$ & 0.0046(7) &&&\\
${^3P}_2E_{yz}(1,loc) - {^1P}_1z(1,loc)$ & 0.0048(8) && 0.0072(40) &\\
${^3P}_2E_{yz}(1,1) - {^1P}_1z(1,1)$ & 0.0039(5) &&&\\
\cline{1-5}
${^1P}_1x(1,loc) - {^3P}_1y(1,loc)$ & 0.0032(4) & 0.0028(6)& 0.0027(38) &
0.0028(43) \\
${^1P}_1x(1,1) - {^3P}_1y(1,1)$ & 0.0026(5) &&&\\
${^1P}_1y(1,loc) - {^3P}_1x(1,loc)$ & 0.0031(3) && 0.0039(42) &\\
${^1P}_1y(1,1) - {^3P}_1x(1,1)$ & 0.0024(4) & & 0.0019(63) &\\
\cline{1-5}
${^1P}_1z(1,loc) -{^3P}_0(1,loc)$ & 0.0097(10) & 0.0097(10)  & 0.0139(49) &
0.013(5) \\
${^1P}_1z(1,1) -{^3P}_0(1,1)$ & & & 0.0120(64) & \\
\end{tabular}
\label{psplits}
\end{table}

\begin{table}
\caption[junk]{$a^{-1}$ values at the  three different values of $\beta$ - the
first error given is statistical, the second, systematic from
higher order relativistic corrections and the third, systematic from
higher order discretisation corrections.
The experimental values for the splittings are 440 MeV ($\overline{\chi_b}
- \Upsilon$) and 563 MeV ($\Upsilon^{'} - \Upsilon$).}
\begin{tabular} {ccccl}
$\beta$ & splitting & $a\Delta M$ & $a\Delta M_g$ & $a^{-1}$ (GeV) \\
\hline
5.7 & $\overline{\chi_b} - \Upsilon$ & 0.326(6) & -0.015 & 1.41(4)(2)(5) \\
& $\Upsilon^{'} - \Upsilon$ & 0.42(4) & -0.007 & 1.36(13)(2)(4) \\
6.0 & $\overline{\chi_b} - \Upsilon$ & 0.174(3) & -0.004 & 2.59(5)(3)(1) \\
& $\Upsilon^{'} - \Upsilon$ & 0.232(8) & -0.002 & 2.45(8)(3)(1) \\
6.2 & $\overline{\chi_b} - \Upsilon$ & 0.127(5) & -0.002 & 3.52(14)(4)(0) \\
& $\Upsilon^{'} - \Upsilon$ & 0.175(8) & -0.001 & 3.24(15)(4)(0) \\
\end{tabular}
\label{scales}
\end{table}

\begin{table}
\caption[junk]{Values for the $\Upsilon$ mass at the three different 
values of $\beta$,
using different prescriptions for $a^{-1}$.
The first error comes from the
statistical error in $aM_{kin}$, the second, the statistical error
in $a^{-1}$ from Table~\ref{scales}.
The experimental value for
the $\Upsilon$ mass is 9.46 GeV. }
\begin{tabular} {ccccc}
$\beta$ & $aM^0_b$ & $a M_{kin}$ & $M_{\Upsilon}$(GeV) & $M_{\Upsilon}$(GeV) \\
&&& $a^{-1}_{(\overline{\chi}_b-\Upsilon)}$ 
\vphantom{$a^{-1}_{\textstyle\mathstrut}$} &
$a^{-1}_{(\Upsilon^{\prime}-\Upsilon)}$ \\
\hline
5.7 & 3.15 & 7.06(7) & 9.95(10)(30) & 9.60(10)(90) \\
6.0 & 1.71 & 3.94(3) & 10.20(8)(20) & 9.65(7)(30) \\
6.2 & 1.22 & 2.89(3) & 10.17(10)(40) & 9.30(10)(40) \\
\end{tabular}
\label{kmass}
\end{table}

\begin{table}
\caption[junk]{Values for the $\Upsilon$ mass in lattice units 
(sixth column) at the three different values of $\beta$,
derived from perturbative renormalisation parameters
$Z_m$ and $E_0$ (see text).}
\begin{tabular} {cccccc}
$\beta$ & $aM^0_b$ & $Z_m$ & $aE_0$ & $aE_{NR}$ & $a(M_{\Upsilon})_{calc}$ \\
\hline
5.7 & 3.15 & 1.25(6) & 0.45(20) & 0.5186(6) & 7.51(55) \\
6.0 & 1.71 & 1.19(4) & 0.30(9) & 0.4537(5) & 3.91(23) \\
6.2 & 1.22 & 1.31(10) & 0.21(4) & 0.3132(3) & 3.09(26) \\
\end{tabular}
\label{bmass}
\end{table}


\begin{references}

\bibitem{improve} G. P. Lepage,
Nucl. Phys. B (Proc. Suppl.) {\bf 60A} (1998) 267.
\bibitem{wittig}
H. Wittig in Proceedings of LAT97, Nucl. Phys.
B (Proc. Suppl.) {\bf 63} 47.
\bibitem{oldups} C.~T.~H.~Davies, K.~Hornbostel, A.~Langnau,
G.~P.~Lepage, A.~Lidsey, J.~Shigemitsu, J.~Sloan, Phys. Rev. D.
 {\bf 50} (1994) 6963.
\bibitem{gpl1} B.~A.~Thacker and G.~P.~Lepage, Phys. Rev. D {\bf 43} (1991)
196.
\bibitem{gpl2}
G. P. Lepage, L. Magnea, C. Nakhleh, U. Magnea, K. Hornbostel,
Phys. Rev. D {\bf 46} (1992) 4052.
\bibitem{japan} NRQCD collaboration, C.~T.~H.~Davies,
 Nucl. Phys. B (Proc. Suppl.)
{\bf 42} (1995) 319;
 C.~T.~H. Davies, 
Nucl. Phys. B (Proc. Suppl.) {\bf 60A} (1998) 124. 
\bibitem{john} J.~Sloan, 
Nucl. Phys. B (Proc. Suppl.) {\bf 60A} (1998) 34. 
\bibitem{evolve} In fact at our smallest lattice spacing ($\beta$ = 6.2)
we use the more symmetrical evolution equation:
\begin{equation}
  G_{t+1} =
\left(1\!-\!\frac{a\delta H}{2}\right)
  \left(1\!-\!\frac{aH_0}{2n}\right)^{n}
 U^\dagger_4
 \left(1\!-\!\frac{aH_0}{2n}\right)^{n}\left(1\!-\!\frac{a\delta H}{2}\right)
G_t
 \quad (t>1) .
\label{pevolve}
\end{equation}
This is similar to the one described in  \cite{gpl2}.
The quark fields undergo a simple rotation from the evolution equation 1, and
so the spectrum is completely unaffected up to $(\delta H)^2$ terms.
\bibitem{tad} G. P. Lepage and P. Mackenzie, Phys. Rev. D {\bf 48} (1993) 2250.
\bibitem{colin} Evidence for this is provided in C. J. Morningstar,
Phys. Rev. D {\bf 48} (1993) 2265.
\bibitem{ukqcd} UKQCD collaboration, R.~D.~Kenway, Nucl. Phys. B
(Proc. Suppl.) {\bf 53} (1997) 209; UKQCD collaboration, H.~P.~Shanahan
{\it et al}, Phys. Rev. D {\bf 55} (1997) 1548.
\bibitem{Kogut} D.~K.~Sinclair, Nucl. Phys. B (Proc. Suppl.)
{\bf 47} (1996) 112.
\bibitem{rdk} R.~D.~Kenway, Proceedings of the XXII International
Conference on High Energy Physics, Leipzig, ed. A.~Meyer and
E.~Wieczorek, Akademie der Wissenschaften der DDR (1984) 51.
\bibitem{noise} G. P. Lepage, in {\it From actions to answers},
World Scientific, 1989.
\bibitem{alpha}
C.~T.~H.~Davies, K.~Hornbostel,
G.~P.~Lepage,  A.~Lidsey, J.~Shigemitsu and J.~Sloan,
Phys.\ Lett.\ {B345}, 42 (1995); Phys. Rev. D{\bf 56} (1997) 2755.
\bibitem{gpllat91} G. P. Lepage, Nucl. Phys. B (Proc.
Suppl.) {\bf 26} (1992) 45.
\bibitem{cambridge} S. M. Catterall, F. R. Devlin,
I. T. Drummond and R. R. Horgan, Phys. Lett. {\bf 321}B
(1994) 246.
\bibitem{bc}
C.~T.~H.~Davies, K.~Hornbostel,
G.~P.~Lepage,  A.~Lidsey, J.~Shigemitsu and J.~Sloan,
Phys. Lett. {\bf 382}B (1996) 131; S. Aoki {\it et al},
Nucl. Phys. B (Proc. Suppl.) {\bf 53} (1997) 355.
\bibitem{colin_again} C. Morningstar, Phys. Rev. D{\bf 50} (1994) 5902;
private communication for results for the evolution equation
used at $\beta$ = 6.2.
\bibitem{us_fb} A. Ali Khan, J. Shigemitsu, S. Collins,
C. T. H. Davies, C. Morningstar, J. Sloan, Phys. Rev. D{\bf 56}
(1997) 7012.
\bibitem{mb}
C.~T.~H.~Davies, K.~Hornbostel, A.~Langnau,
G.~P.~Lepage,  A.~Lidsey, J.~Shigemitsu and J.~Sloan,
Phys. Rev. Lett. {\bf 73} (1994) 2654.
\bibitem{mass_in_prep} NRQCD collaboration, in preparation.
\bibitem{rowland} P. Rowland, PhD thesis, University
of Edinburgh, 1997.
\bibitem{gf11} F. Butler, H. Chen. J. Sexton,
A. Vaccarino and D. Weingarten,
Nucl. Phys. B{\bf 430} (1994) 179. We use their
results at $\beta$ = 5.7 and 6.17 to compare to
our results at $\beta$ = 5.7 and 6.2.
\bibitem{manke} T. Manke, R. Horgan, I.T. Drummond
and H. P. Shanahan,
Phys. Lett. B{\bf 408} (1997) 308; T. Manke, Proceedings of
LAT97 conference, Edinburgh 1997, Nucl. Phys.
B (Proc. Suppl.) {\bf 63} 332.
\bibitem{a_note} Note that in previous papers,  \cite{alpha}
where plots of the fine structure have been shown we have used
an average $a^{-1}$ from two different splittings to set the scale,
and taken results from only a single value of $\beta$.
An analysis of scaling is much clearer if only a single splitting
is used for $a^{-1}$ at all values of $\beta$ and this is
why results in physical
units here will differ slightly from those of \cite{alpha}.
\bibitem{charm}
C.~T.~H.~Davies, K.~Hornbostel,
G.~P.~Lepage,  A.~Lidsey, J.~Shigemitsu and J.~Sloan,
Phys. Rev. D{\bf 52} (1995) 6519.
\bibitem{v6_note} In fact in their higher order action
Manke {\it et al} also included an improved
derivative and field in
the spin-independent Darwin term, i.e. some of the
spin-independent $\cal{O}$$(a^2v^4)$ terms. As described
in the analysis of systematic errors, this should not
give a significant effect to $a^{-1}$.
\bibitem{trottier} H. Trottier,
Phys. Rev. D{\bf 55} (1997) 6844.
\bibitem{trottier2} G. P. Lepage and H. Trottier in
Proceedings of LAT97 conference, Edinburgh 1997,
Nucl. Phys. B (Proc. Suppl.) {\bf 63} 865.
\bibitem{c4_note} In fact the results for $c_4^{(1)}$ in
the above reference do not match exactly in $Ma$ values
to the ones that we have used here. We just take the
nearest result.
\bibitem{SESAM} N. Eicker {\it et al}, Phys.
Rev. D{\bf 57} (1998) 4080;
A. Spitz, in
Proceedings of LAT97 conference, Edinburgh 1997,
Nucl. Phys. B (Proc. Suppl.) {\bf 63} 317.
\bibitem{schladming} See, for example, the discussion
in C. Davies, lectures given at the Schladming winter school
1997, hep-ph/9710394.
\bibitem{jonivar} We thank Jon Ivar Skullerud (UKQCD) for
values of $u_{0L}$ at $\beta$ = 6.0 and 6.2. We use
$u_{0L}$ = 0.824 at $\beta$ = 5.7, 0.861 at $\beta$
= 6.0 and 0.857 at $\beta$ = 6.2.
\bibitem{beth} B. A. Thacker and C. T. H. Davies,
Phys. Rev. D{\bf 48} (1993) 1329; G. Bodwin, D. Sinclair
and S. Kim, Phys. Rev. Lett. {\bf 77} (1996) 2376.
\end{references}
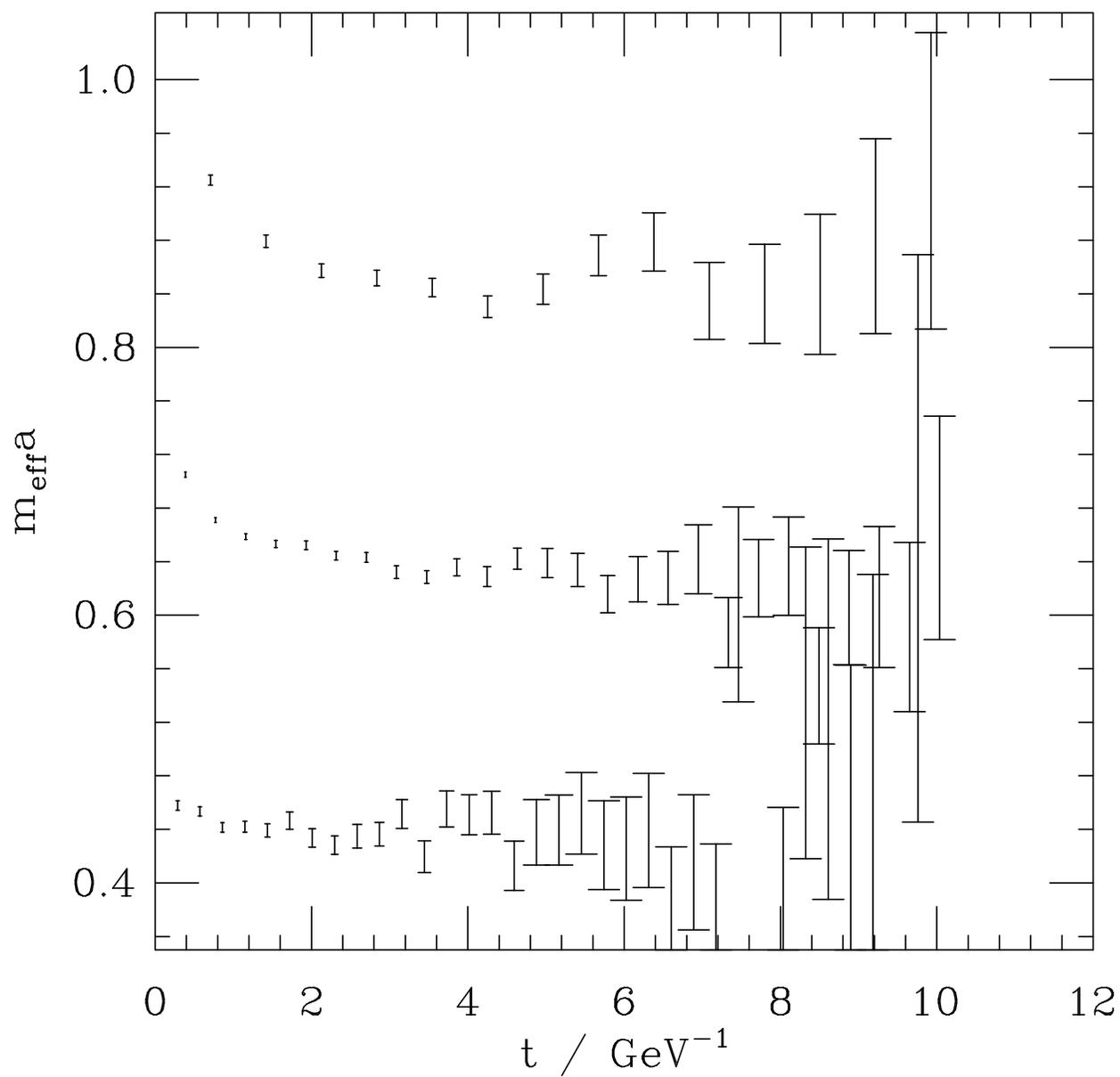
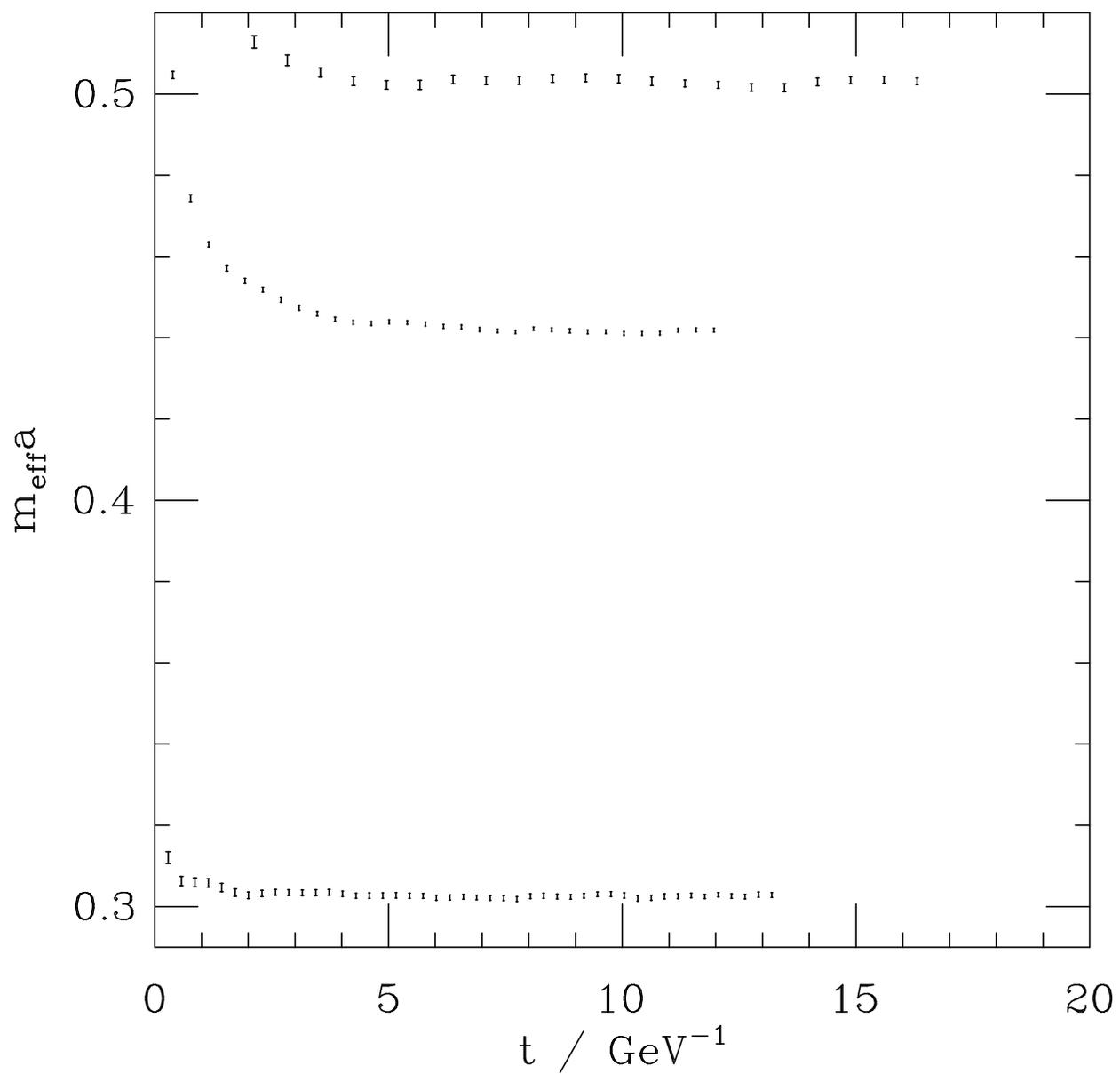
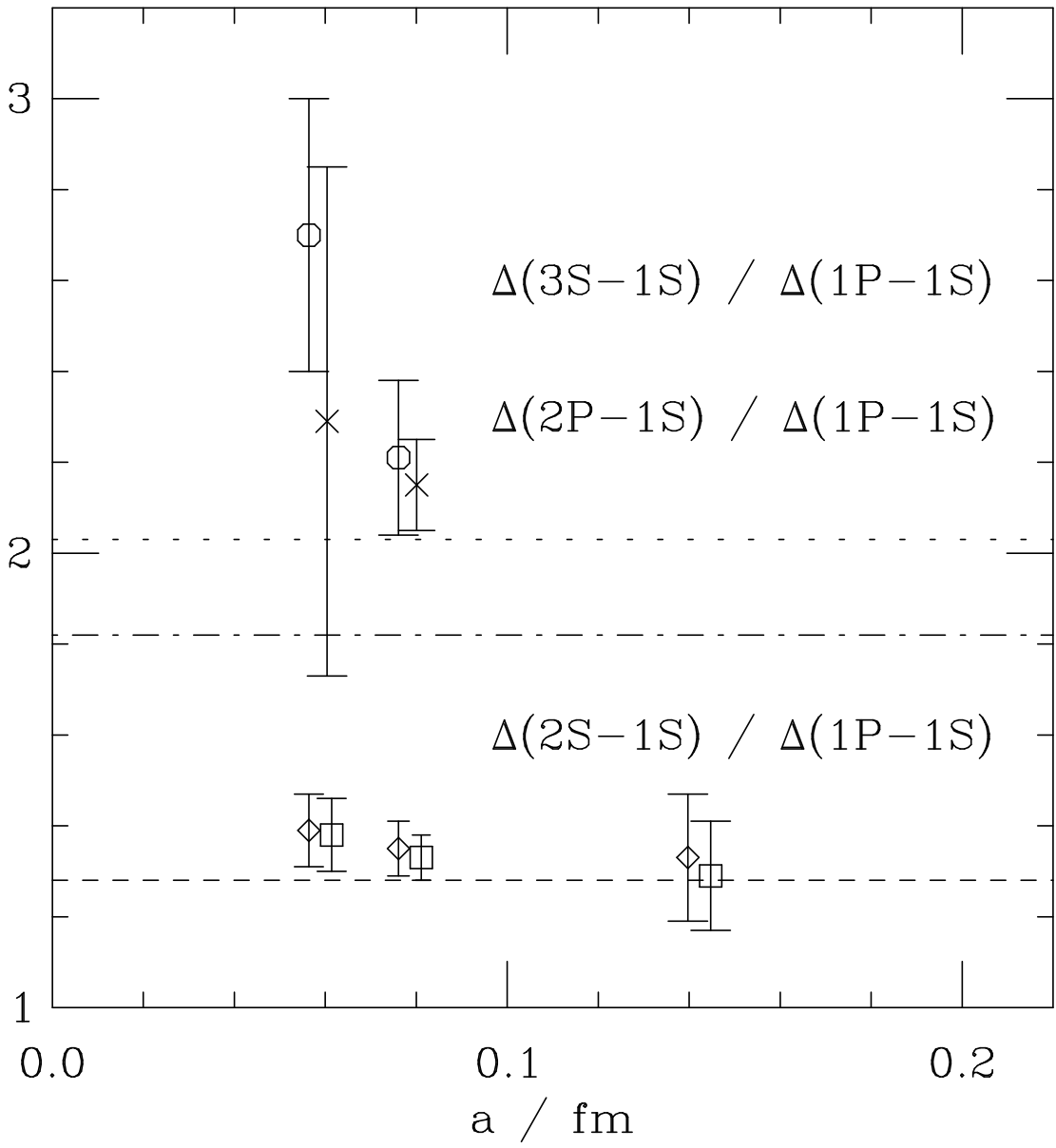
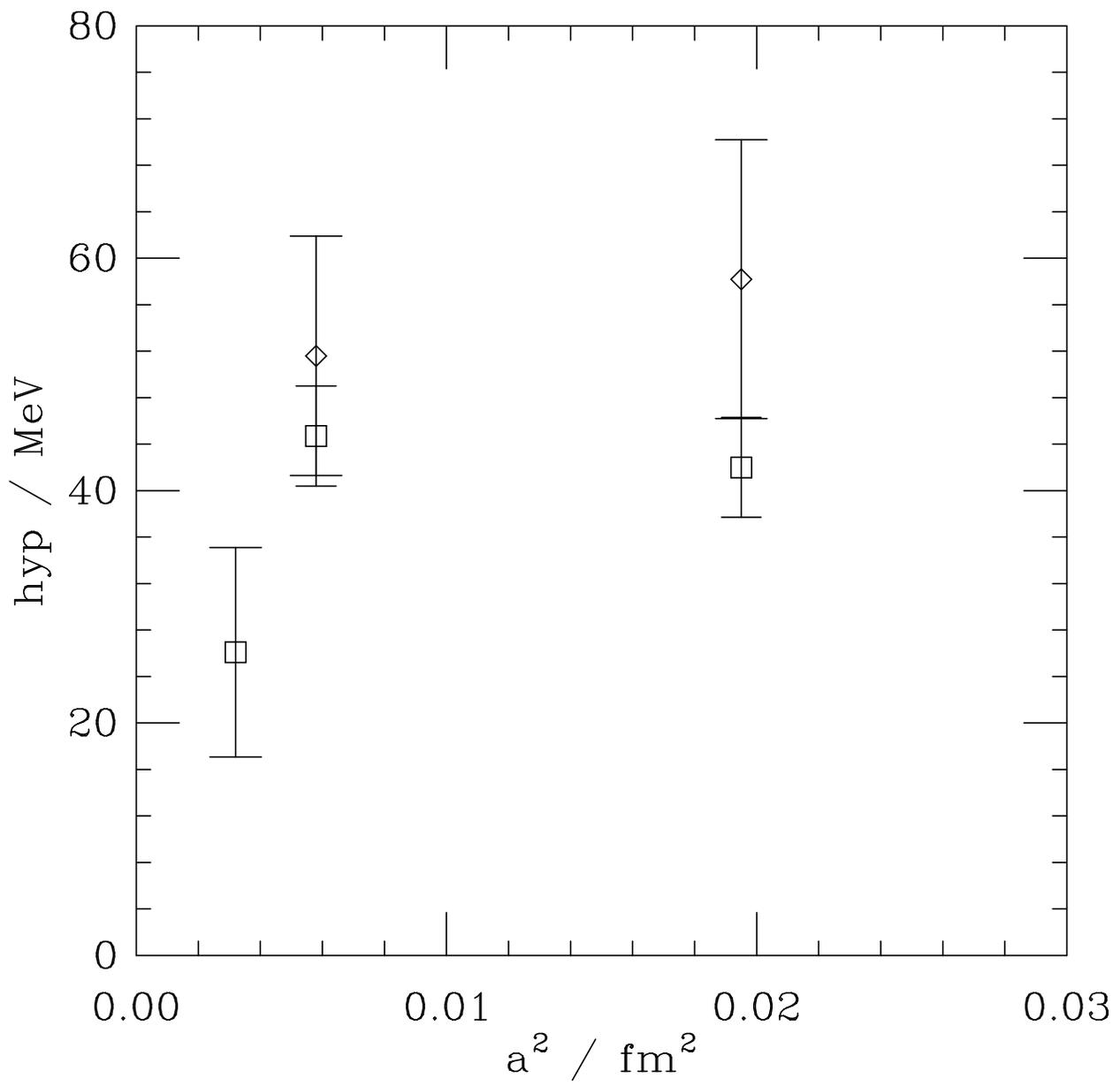
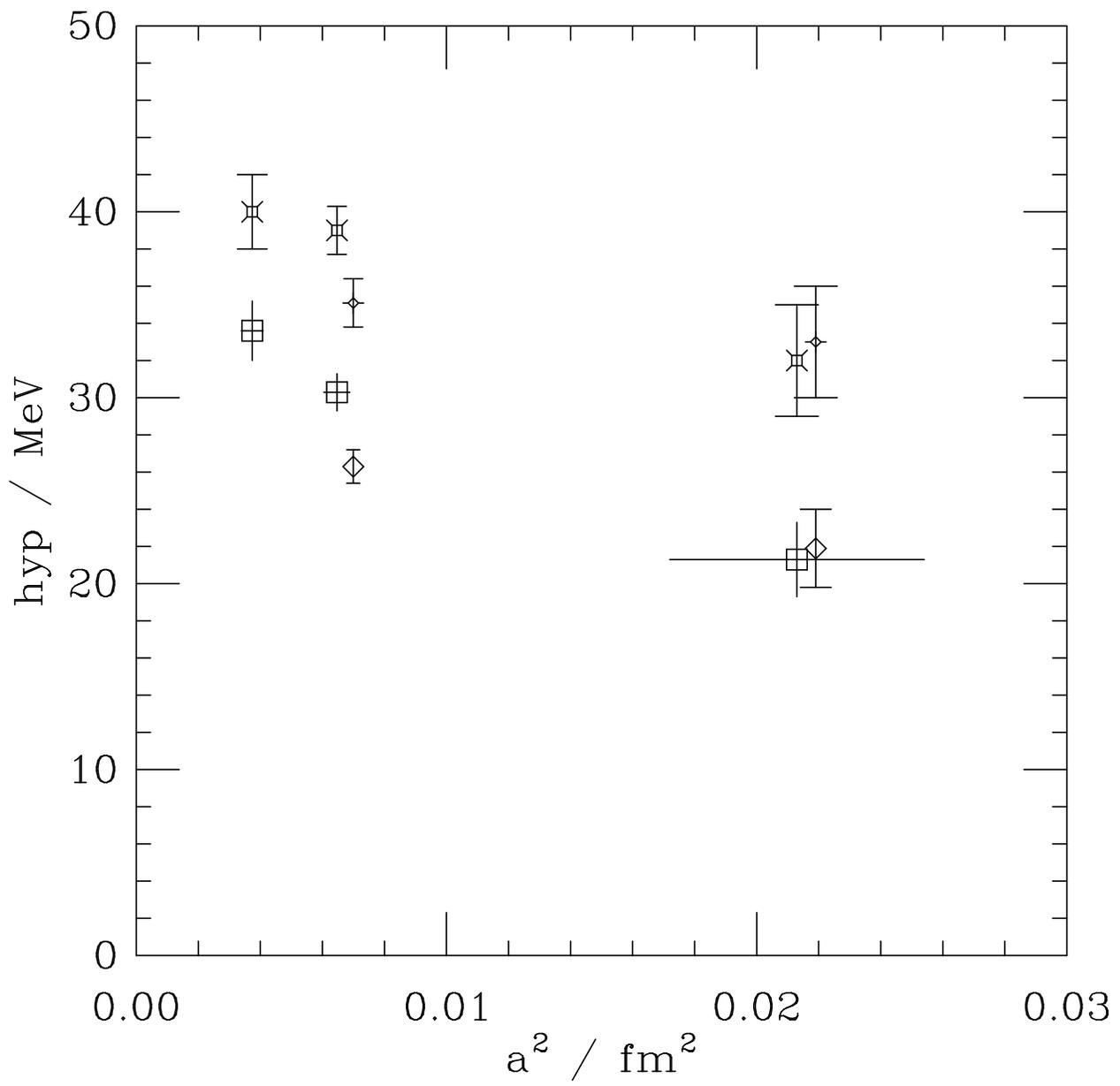
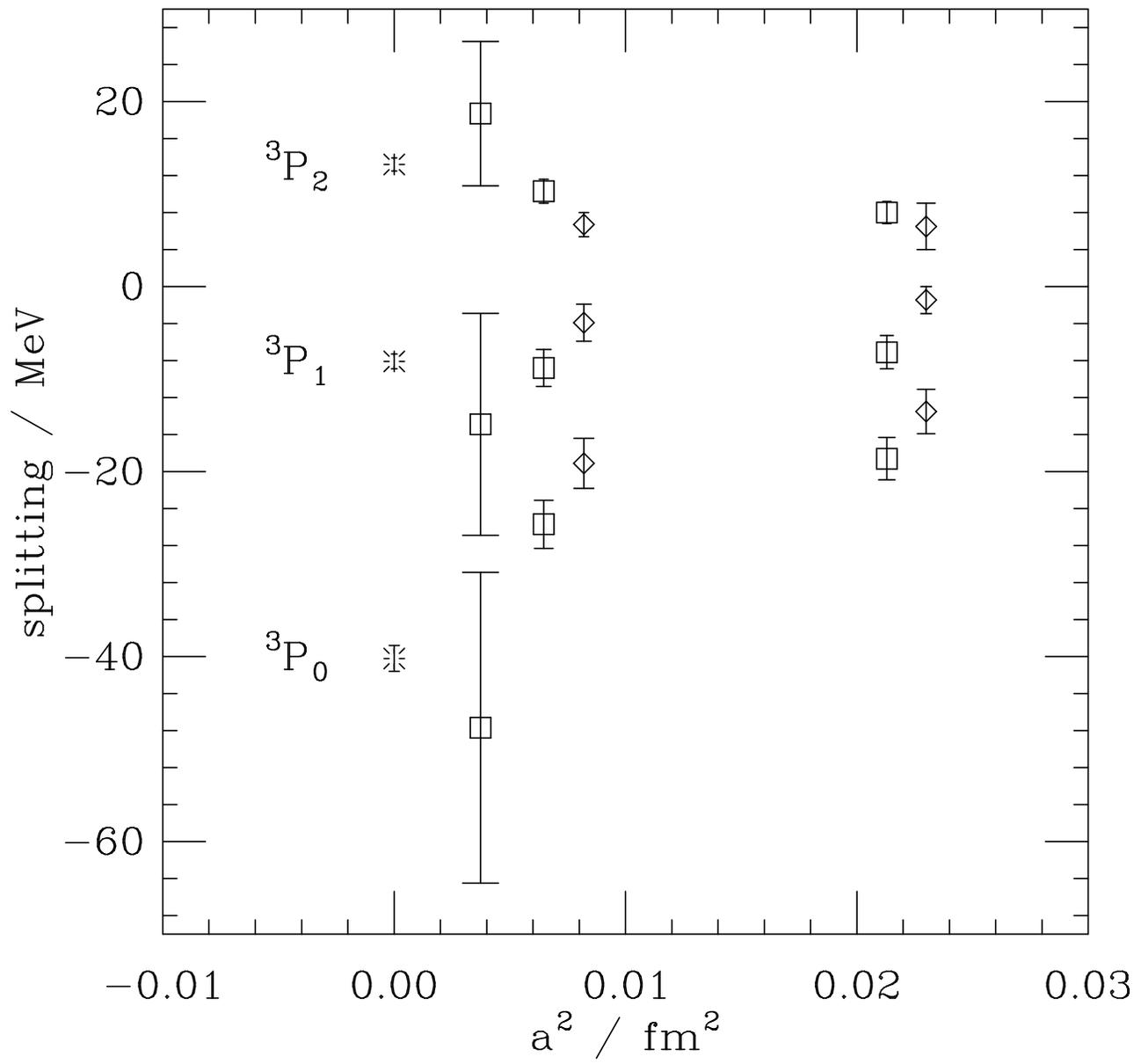
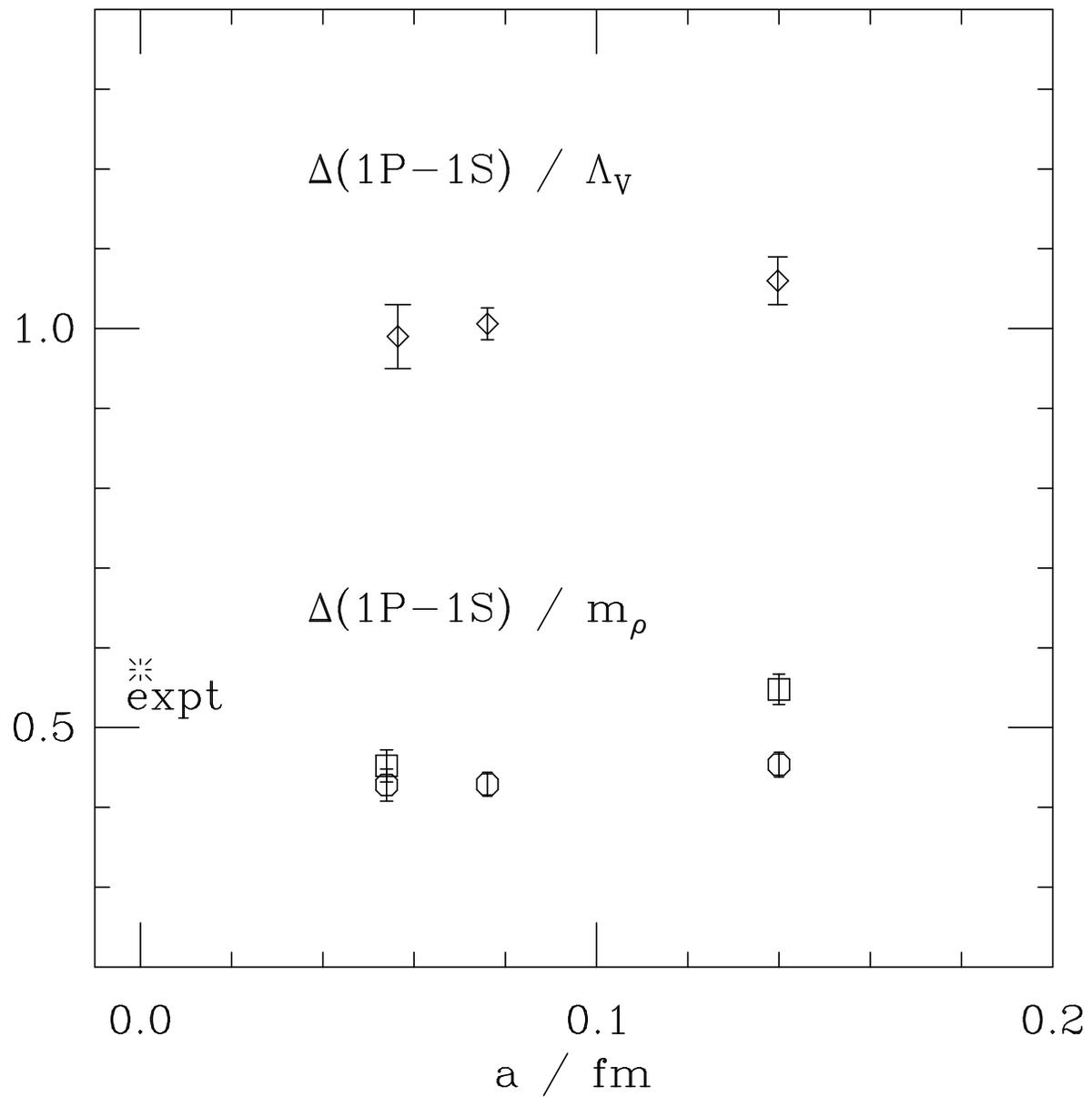
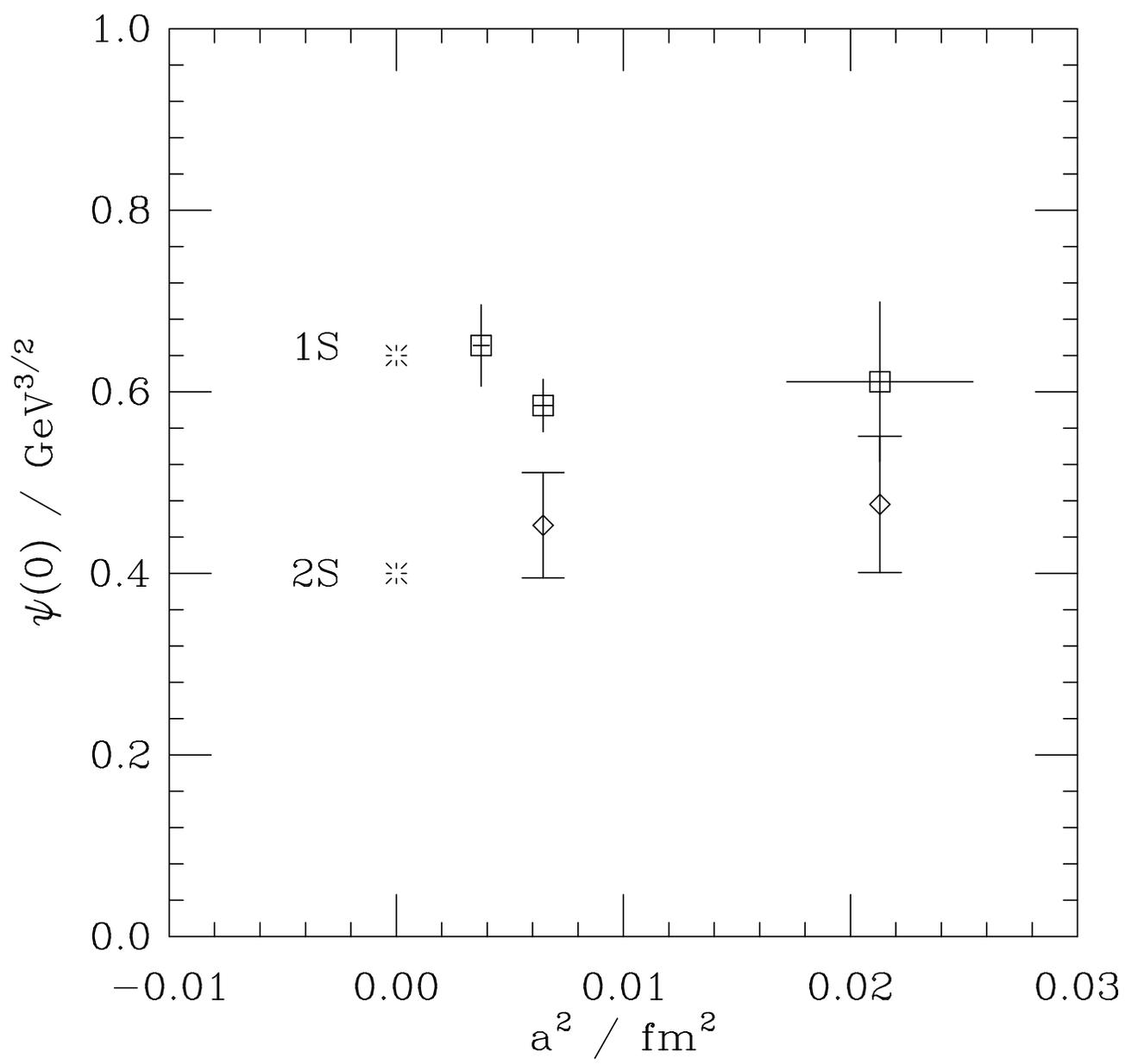
\end{document}